\documentclass[english]{iopart}

\usepackage{epsfig}
\usepackage[T1]{fontenc}
\usepackage[latin1]{inputenc}
\usepackage{graphicx}

\begin{document}

\title[Singularities of $\, n$-fold integrals]{\Large
Singularities of $\, n$-fold integrals of the Ising class
and the theory of elliptic curves}

\author{ 
S. Boukraa$^\dag$, S. Hassani$^\S$, 
J.-M. Maillard$^\ddag$ and N. Zenine$^\S$ }
\address{\dag LPTHIRM and D\'epartement d'A{\'e}ronautique,
 Universit\'e de Blida, Algeria}
\address{\S  Centre de Recherche Nucl\'eaire d'Alger, \\
2 Bd. Frantz Fanon, BP 399, 16000 Alger, Algeria}
\address{\ddag\ LPTMC, Universit\'e de Paris 6, Tour 24,
 4\`eme \'etage, case 121, \\
 4 Place Jussieu, 75252 Paris Cedex 05, France} 
\ead{maillard@lptmc.jussieu.fr, maillard@lptl.jussieu.fr, 
boukraa@mail.univ-blida.dz, njzenine@yahoo.com}

\begin{abstract}
We introduce some  multiple integrals 
that are expected to have the  same singularities as the 
singularities of the $\, n$-particle contributions $\chi^{(n)}$
to the susceptibility of the square lattice Ising model. 
We find the Fuchsian linear differential equation satisfied by these 
 multiple integrals for $\, n=1, 2, 3, 4$ and
only modulo some primes for $\, n=5$ and $\, 6$, thus providing a large set
of (possible) new singularities of the $\chi^{(n)}$. 
We discuss the singularity structure for these multiple integrals
by solving the Landau conditions.
We find that the singularities of the associated ODEs
identify (up to $n=\, 6$) with the leading pinch Landau
singularities. The second remarkable obtained feature
is that the singularities of the ODEs associated with
the multiple integrals reduce to the singularities of
the ODEs associated with a {\em finite number of one
dimensional integrals}. Among the singularities found, 
we underline the fact that the quadratic  polynomial condition
$ 1+3\, w \, +4 \, w^2\, = \, 0$, that occurs in
the  linear differential equation of $\, \chi^{(3)}$, 
actually corresponds to a remarkable property of selected   
elliptic curves, namely the occurrence
of complex multiplication. The interpretation of complex multiplication
 for elliptic curves as complex fixed points of the selected
 generators of the renormalization group, namely isogenies of elliptic curves,
is sketched. Most of the other singularities occurring in our multiple integrals 
are not related to complex multiplication situations, suggesting 
an interpretation in terms of (motivic) mathematical structures
beyond the theory of elliptic curves.

\end{abstract} 
\vskip .5cm


\noindent {\bf AMS Classification scheme numbers}: 34M55, 
47E05, 81Qxx, 32G34, 34Lxx, 34Mxx, 14Kxx 


\section{Introduction}
\label{recalls}
The susceptibility $\, \chi$ of the square lattice Ising model
has been shown by  Wu, McCoy, Tracy and
 Barouch~\cite{wu-mc-tr-ba-76} to be expressible as an infinite 
sum of holomorphic functions,  
given as multiple 
integrals, denoted $\, \chi^{(n)}$, that is 
$\, kT \cdot \chi\, = \, \, \sum \,  \chi^{(n)}$.
B. Nickel found~\cite{nickel-99, nickel-00} 
that each of these $\, \chi^{(n)}$'s
is actually singular on a set of points located on the unit circle
$\, |s| \, = \, |\sinh(2\, K)| \, = \,  1$, where 
$K=\, J/kT$ is the usual Ising model temperature variable.

These singularities are located at solution points of the following equations:
 \begin{eqnarray}
\label{sols}
&& \, 2 \cdot 
\Bigl(s \, + \, {{1} \over {s}}\Bigr) \, = \, \, \, \, 
u^k\, + \,{{1} \over {u^k}} \,
+ \, u^m\, + \,{{1} \over {u^m}} \,\nonumber \\
&&  \qquad u^{2\, n+1} \, = \, \, 1,  \qquad   \qquad  
-n \,  \,\le\,  \, m, \, \, k\,  \,\,\le \, \, n  
\end{eqnarray}
From now on, we will call these singularities of the ``Nickelian type'', 
or simply ``Nickelian singularities''.
The accumulation of this infinite set of singularities of the higher-particle
components of $\chi(s)$ on the unit circle $\, |s|\, = \, 1$,
leads, in the absence of mutual cancellation, to some consequences regarding the
non holonomic (non D-finite) character of the susceptibility,
possibly building a natural boundary for the total $\,\chi(s)$.
However, it should be noted that
 new singularities that are not of the ``Nickelian type''
were discovered as singularities of the Fuchsian linear differential equation
associated~\cite{ze-bo-ha-ma-04, ze-bo-ha-ma-05, ze-bo-ha-ma-05c} with
 $\chi^{(3)}$ and as singularities of  $\chi^{(3)}$ {\em itself}~\cite{bo-ha-ma-ze-07}
 but seen as a function of $ \, s$.
They correspond to the quadratic  polynomial
$\, 1\,+3\, w \, +4 \, w^2$ where $2\,w=\, s/(1+s^2)$.
In contrast with this situation, the  Fuchsian linear differential equation,
associated~\cite{ze-bo-ha-ma-05b} with  $\, \chi^{(4)}$, does not provide any 
new singularities. 

Some remarkable Russian-doll structure  as well as direct sum
decompositions  were found for the corresponding linear differential
operators for $\, \chi^{(3)}$ and $\, \chi^{(4)}$.
 In order to understand the ``true nature''
of the susceptibility of the square lattice Ising model, it is of fundamental
importance to have a better understanding of the singularity structure
of the $n$-particle contributions $\chi^{(n)}$, and also of 
the mathematical  structures associated with these
$\, \chi^{(n)}$, namely the {\em infinite set} of (probably Fuchsian)
linear differential equations associated with this infinite set of holonomic
functions.
Finding more Fuchsian linear differential equations having the $\chi^{(n)}$'s
as solutions, beyond those already found~\cite{ze-bo-ha-ma-04, ze-bo-ha-ma-05b}
for $\, \chi^{(3)}$ and $\, \chi^{(4)}$, probably requires the performance
of a large set of analytical, mathematical and
computer programming ``tours-de-force''. 

 As an alternative, and in order to bypass this ``temporary''
obstruction, we have developed, in parallel, a new strategy.

We have introduced~\cite{bo-ha-ma-ze-07}
some  single (or multiple) ``model'' integrals
as an ``ersatz'' for the $\, \chi^{(n)}$'s as far as the locus
of the singularities is concerned. 
The $\, \chi^{(n)}$'s are defined by 
 $(n-1)$-dimensional integrals~\cite{nickel-00,pal-tra-81,yamada-84}
 (omitting the prefactor\footnote[2]{The
prefactor reads $\, (1-s^4)^{1/4}/s$ for 
$\, T \, > \, T_c$ and $\, (1-s^{-4})^{1/4}$ 
for $T \, < \, T_c$. })
\begin{eqnarray}
\label{chi3tild}
\tilde{\chi}^{(n)}\, =\,\,\, {\frac{(2 w)^n}{n!}}  
\prod_{j=1}^{n-1}\int_0^{2\pi} {\frac{d\phi_j}{2\pi}}
\Bigl( \prod_{j=1}^{n} y_j \Bigr) \cdot   R^{(n)}
\cdot  \Bigl( G^{(n)} \Bigr)^2
\end{eqnarray}
where
\begin{eqnarray}
\label{Gn}
G^{(n)}\,=\,\,\,
\Bigl( \prod_{j=1}^{n} x_j \Bigr)^{(n-1)/2}
 \,\,\prod_{1\le i < j \le n}
{\frac{2\sin{((\phi_i-\phi_j)/2)}}{1-x_ix_j}}
\end{eqnarray}
and
\begin{eqnarray}
R^{(n)} \, = \, \,\,
 {\frac{1+\prod_{i=1}^{n}x_i}{1-\prod_{i=1}^{n}x_i}}
\end{eqnarray}
with
\begin{eqnarray}
\label{thex}
&&x_{i}\, =\,\,\,  \,  \frac{2w}{1-2w\cos (\phi _{i})\, 
+\sqrt{\left( 1-2w\cos (\phi_{i})\right)^{2}-4w^{2}}},    \\
\label{they}
&&y_{i} \, = \, \, \,
\frac{1}{\sqrt{\left(1\, -2 w\cos (\phi _{i})\right)^{2}\, -4w^{2}}}, 
\quad
 \quad  \quad  \quad \sum_{j=1}^n \phi_j=\, 0.  
\end{eqnarray}

The two families of integrals we have
 considered in~\cite{bo-ha-ma-ze-07}
are very rough approximations of the integrals (\ref{chi3tild}). 
For the first family\footnote[1]{Denoted $\, Y^{(n)}(w)$
 in~\cite{bo-ha-ma-ze-07}.}, we considered the $\, n$-fold
integrals corresponding to the 
 product of (the square\footnote[2]{Surprisingly the integrand with 
$\, ( \prod_{j=1}^{n} y_j )^2$ yields
second order linear differential equations~\cite{bo-ha-ma-ze-07}, and
 consequently, we have been able
to totally decipher the corresponding singularity structure.
By way of contrast the integrand with the simple product
$\, ( \prod_{j=1}^{n} y_j )$ yields
 linear differential equations of higher order, but with 
identical singularities~\cite{bo-ha-ma-ze-07}.}
 of the) $y_i$'s,
integrated over the whole domain of integration of the
$\phi_i$ (thus getting rid of the factors
$\, G^{(n)}$ and $\, R^{(n)}$). Here, we found a
 subset of singularities occurring in 
the $\chi^{(n)}$ as well as the quadratic 
polynomial condition $\, 1\, +3w\, +4w^2\,  =\, \, 0$.

For the second family, we discarded the factor
$\, G^{(n)}$ and the product of $\, y_i$'s, and we restricted the domain of
integration to the principal diagonal of the angles $\phi_i$
($\phi_1\, = \,\phi_2\, = \,\cdots \, = \, \phi_{n-1}$).
These simple integrals (over a {\em single} variable), were
denoted~\cite{bo-ha-ma-ze-07} $\, \Phi_{D}^{(n)}$:
\begin{eqnarray}
\label{chinaked}
\Phi_{D}^{(n)}\,=\,\,\,\,  
 -{{1} \over {n!}}\,\,  \,+ {{2} \over {n!}} \, \int_0^{2\pi} 
{\frac{d\phi}{2\pi}}
 \,   \,   {\frac{1}{1\, -x^{n-1}(\phi)  \cdot  x ((n-1)\phi)}} 
\end{eqnarray}
where $\, x(\phi)$ is given by (\ref{thex}). 

Remarkably these very simple integrals both {\em reproduce 
all the singularities}, discussed by
Nickel~\cite{nickel-99,nickel-00}, as well as
the quadratic roots of $\, 1\,+3w\,+4w^2\,=\,0$ 
found~\cite{ze-bo-ha-ma-04,ze-bo-ha-ma-05}
 for the linear ODE of $\, \chi^{(3)}$.
One should however note that, in contrast with the
$\, \chi^{(n)}$, no Russian-doll
or direct sum decomposition structure is found for the  
linear differential operators corresponding 
to these $\,\Phi_{D}^{(n)}$. 

Another approach has been introduced 
as a simplification of the susceptibility of the Ising model
 by  considering a magnetic field restricted to one diagonal of the
square lattice~\cite{Diag}. For this ``diagonal susceptibility'' 
model~\cite{Diag}, we benefited from
the {\em form factor decomposition} of the diagonal two-point correlations 
$\, C(N,N)$, that has been recently 
presented~\cite{Holo}, and subsequently proved by
 Lyberg and  McCoy~\cite{lyb-mcc-07}.
The corresponding $\, n$-fold integrals  $\, \chi^{(n)}_d$
were found to exhibit remarkable direct sum structures inherited
from the  direct sum structures of the  form factor~\cite{Diag,Holo}.
The linear differential operators of the form factor~\cite{Holo}
being closely linked to the second order differential operator $\, L_E$ (resp. $\, L_K$)
of the complete elliptic integrals $\, E$ (resp. $\, K$),
this ``diagonal susceptibility'' model~\cite{Diag} is closely linked to the
elliptic curves of the two-dimensional Ising model.
By way of contrast, we note that the singularities of the linear ODE's for 
these $\, n$-fold integrals~\cite{Diag}  $\, \chi^{(n)}_d$
are quite elementary (consisting of only $n$-th roots of unity) in comparison 
with the singularities we encounter  for the integrals 
on a single variable (\ref{chinaked}).

These two approaches  corresponding to two
different sets of $\, n$-fold integrals of the Ising class~\cite{crandall} 
are complementary:  (\ref{chinaked}) is more dedicated to reproduce 
the non-trivial head polynomials encoding the location of the singularities
of the $\, \chi^{(n)}$, but fails to reproduce some remarkable 
(Russian-doll, direct sum decomposition) algebraico-differential structures of the 
 corresponding linear differential operators, while the other one~\cite{Diag} 
preserves these non-trivial  structures of the 
 corresponding linear differential operators but provides a poorer representation of 
the location of the singularities ($n$-th roots of unity). 

\vskip .1cm 

In this paper, we return to the integrals (\ref{chi3tild})
where, this time, the natural next step is 
to consider the following family of $\, n$-fold integrals
\begin{eqnarray}
\label{In}
\Phi_H^{(n)} \, \,= \,\,\,\,  {\frac{1}{n!}}  \cdot 
 \prod_{j=1}^{n-1} \int_0^{2\pi} {\frac{d\phi_j}{2\pi}} \cdot 
\Bigl( \prod_{j=1}^{n} y_j \Bigr)  \cdot  
 {\frac{1\,+\prod_{i=1}^{n}\, x_i}{1\,-\prod_{i=1}^{n}\, x_i}} 
\end{eqnarray}
which amounts to  getting rid of the (fermionic) factor $\, (G^{(n)})^2$
 in the $\, n$-fold integral (\ref{chi3tild}).
 This family is as close as possible to 
(\ref{chi3tild}), for which we know that
finding the corresponding linear differential ODE's
is a huge task.
The idea here is that the methods and techniques we have 
developed~\cite{ze-bo-ha-ma-04,ze-bo-ha-ma-05}
for series expansions calculations of $\, \chi^{(3)}$
and $\, \chi^{(4)}$, seem to indicate that the
quite involved fermionic term $\, (G^{(n)})^2$
in the integrand of (\ref{chi3tild})
should not impact greatly on the location of singularities
 of these $\, n$-fold integrals  (\ref{chi3tild}).
This is the best simplification of the integrand of (\ref{chi3tild})
for which we can expect to retain much exact information about the
location of the singularities of the original Ising problem. 
However, we certainly do not expect to recover from 
the $\, n$-fold integrals (\ref{In})
the local singular behavior (exponents, 
amplitudes of singularities, etc ...).
Getting rid of the (fermionic) factor $\, (G^{(n)})^2$
are we moving away from the 
elliptic curves of the two-dimensional Ising model ?
Could it be possible that we lose the strong 
(Russian-doll, direct sum decomposition) 
algebraico-differential structures of the 
 corresponding linear differential operators inherited from  
the second order differential operator $\, L_E$ (resp. $\, L_K$)
of the complete elliptic integrals $\, E$ (resp. $\, K$)
but keep some characterization of elliptic curves 
through more  ``primitive'' (universal) features of these
$\, n$-fold integral like the location of their singularities ?

In the sequel, we give the expressions of $\, \Phi_H^{(1)}$,
  $\, \Phi_H^{(2)}$ and the Fuchsian
linear differential equations for  
$\, \Phi_H^{(n)}$ for $\, n\,=\, 3$ and $\, n\,=\, 4$.
For $\,n\,=\, 5,\,  6$, the computation 
(linear ODE search of a series) becomes
much harder. Consequently we use a 
{\em modulo prime} method to obtain the form of the
corresponding linear ODE with totally explicit singularity structure.
These results provide a large set of ``candidate singularities'' for
the $\, \chi^{(n)}$.
From the resolution of the Landau 
conditions~\cite{bo-ha-ma-ze-07} for (\ref{In}),
we show that the singularities of (the linear ODEs of) 
these multiple integrals actually reduce to the concatanation
of the singularities of (the linear ODEs of) 
 a set of one-dimensional integrals.
We discuss the {\em mathematical, as well as physical,
interpretation} of these new singularities.
In particular we will see that they correspond to {\em pinched 
Landau-like singularities} as previously 
noticed by  Nickel~\cite{nickel-05}.
Among all these polynomial singularities, the
 quadratic numbers $\, 1\, +3\, w\, +4\, w^2\, =\, 0$ are
highly selected. We will show that these selected 
quadratic numbers are related to
{\em complex multiplication for the elliptic 
curves} parameterizing the square Ising model.

The paper is organized as follows.
Section (\ref{singODE})
presents the multidimensional integrals $\,\Phi_H^{(n)}$ and
the singularities of the corresponding linear
 ODE for $\, n=\,3,\, \cdots,\, 6\,$,
that we compare with the singularities 
obtained from the Landau conditions. We show that the 
set of singularities associated with the ODEs 
of the multiple integrals $\, \Phi^{(n)}_H$ reduce to the singularities of
the ODEs associated with a {\em finite number of one-dimensional integrals}.
Section (\ref{bridge}) deals with the {\em complex
 multiplication for the elliptic 
curves} related to the  singularities given
 by the zeros of the quadratic polynomial
$\, 1\, +3w\, +4\, w^2\, =\, 0$.
Our conclusions are given in section (\ref{conclu}).

\section{The singularities of the linear ODE for $\, \Phi_H^{(n)}$}
\label{singODE}

For the first two values of $n$, one obtains
\begin{eqnarray}
\Phi_H^{(1)} \, = \,\,\,  \quad{\frac{1}{1-4w}}
\end{eqnarray}
and
\begin{eqnarray}
\Phi_H^{(2)} \, = \,\,\,\, {{1} \over {2}} \cdot {\frac{1}{1-16w^2}} \cdot 
{_2}F_1 ( 1/2, -1/2; 1; 16w^2). 
\end{eqnarray}

For $n \ge 3$, the  series  coefficients of the multiple 
integrals $\,\Phi_H^{(n)}$ are obtained by expanding in the variables
$x_i$ and performing the integration (see \ref{a0}). One obtains
\begin{eqnarray}
\label{integratedsum}
\Phi_H^{(n)}\, =\, \,\, 
 {\frac{1}{n!}}  \cdot  \sum_{k=0}^{\infty} \sum_{p=0}^{\infty}\,
(2\, -\delta_{k,0}) \cdot (2\, -\delta_{p,0}) \cdot  w^{n(k+p)} \cdot a^n(k,p)
\end{eqnarray}
where $\, a(k,p)$ is a $\, \, \, _4\,F_3$ hypergeometric
 series dependent on $\,w$.

The advantage of using these simplified integrals
 (\ref{In}) instead of the original
ones (\ref{chi3tild}) is twofold. 

Using (\ref{integratedsum}) the series generation
 is straightforward compared to
the complexity related to the $\chi^{(n)}$. 
As an illustration  note that on a desk
computer, $\, \Phi_H^{(n)}$ are generated up to $\,  w^{200}$
 in less than 10 seconds CPU time
for all values of $\, n$,
while the simplest case of the $ \, \chi^{(n)}$, namely  $ \, \chi^{(3)}$,
 tooks three minutes to generate the series up to $\,  w^{200}$.
This difference between
the $\, \Phi_H^{(n)}$ and  $ \, \chi^{(n)}$ increases rapidly with
increasing $\, n$ and increasing number of generated terms.
We note that for the
$\, \Phi_H^{(n)}$ quantities and for a fixed order, the CPU time
is decreasing\footnote[8]{This can be seen from the series expansion (\ref{integratedsum}).
Denoting $\, R_0$ the fixed order, one has $\, n \cdot (p+k)\, \le \, R_0$, while
the CPU time for the series generation of $\,a^n(k,\, p)$
 is not strongly dependent on $\, n$.} with increasing $\, n$. For  $ \, \chi^{(n)}$
the opposite is the case.

The second point is that,
for a given $n$, the linear ODE can be found with less terms in the series
compared to the linear ODE for the $\chi^{(n)}$. Indeed for  $\chi^{(3)}$,  
360 terms were needed while 150
terms were enough for $\, \Phi_H^{(3)}$. The same feature holds
for $\chi^{(4)}$ and  $\, \Phi_H^{(4)}$ (185 terms for $\chi^{(4)}$
and 56 terms\footnote[5]{From now on, for even $\, n$, the number of terms stands
for the number of terms in the variable $x=w^2$.}
 for $\, \Phi_H^{(4)}$).

With the fully integrated sum (\ref{integratedsum}),
 a sufficient number of terms is generated to obtain the linear
 differential equations.
We succeeded in obtaining  the linear differential equations, respectively
of minimal order five and six, corresponding to
$\,\Phi_H^{(3)}$ and $\,\Phi_H^{(4)}$. These linear ODE's
 are given in \ref{a}.

For  $\,\Phi_H^{(n)}$  ($n\, \ge \, 5$), the calculations, in order to get
the linear ODEs become really huge\footnote[2]{Except
the  generation of large series which remains reasonable.}.
For this reason, we introduce a modular strategy which
amounts to generating large series {\em modulo a prime} and
then deducing the ODE modulo that prime.
Note that the ODE of minimal order is {\em not necessarily
the simplest one} as far as the required number of terms
in the series expansion to find the linear ODE is concerned.
We have already encountered such a situation~\cite{ze-bo-ha-ma-05b,Diag}.
For $\,\Phi_H^{(5)}$ (resp. $\,\Phi_H^{(6)}$), the
 linear ODE of  minimal order is
of order 17 (resp. 27) and needs 8471 (resp. 9272)
terms in the series expansion to be found.

Actually, for $\,\Phi_H^{(5)}$ (resp. $\,\Phi_H^{(6)}$), we have found
the corresponding linear ODEs of order 28 (resp. 42) with {\em only}
2208 (resp. 1838) terms from which we have deduced the
minimal ones.

The form of these two minimal order linear ODEs obtained
modulo primes is sketched in \ref{a}.
In particular, the singularities (given by the roots
 of the head polynomial in front of the highest
order derivative), are given with the corresponding multiplicity  in \ref{a}.
Some details about the ODE search are also given in \ref{a}.

We have also obtained  very long series (20000 coefficients) modulo
primes  for $\,\Phi_H^{(7)}$, but, unfortunately, this has not 
been sufficient to identify the linear ODE (mod. prime) up to order 100.

The singularities of the linear ODE for the first $\,\Phi_H^{(n)}$ 
 are respectively zeros of the following polynomials (besides $w=\infty$):
\begin{eqnarray}
\label{singphi5}
n=3, \quad  && w \cdot  \left( 1-16\,w^2 \right)
  \left( 1-w \right)
\left( 1+2\,w \right)  \left( 1+3\,w+4\,{w}^{2} \right), \nonumber \\
n=4,\quad  && w  \cdot   \left( 1-16\,w^2 \right) 
 \left(1-4\,w^2 \right), \nonumber \\
\label{singphi5b}
n=5, \quad  && w  \cdot   \left( 1-16\,w^2 \right)
  \left( 1-w^2 \right)
\left( 1+2\,w \right) \left( 1+3\,w+4\,{w}^{2} \right) \nonumber \\
&&  \left( 1-3\,w+{w}^{2} \right)  \left( 1+2\,w-4\,{w}^{2}
 \right)  \left( 1+4\,w+8\,{w}^{2} \right) \nonumber \\
&&  \left( 1-7\,w+5\,{w}^{2}-4\,{w}^{3} \right) 
 \left( 1-w-3\,{w}^{2}+4\,{w}^{3} \right) \nonumber \\
&& \left( 1+8\,w+20\,{w}^{2}+15\,{w}^{3}+4\,{w}^{4} \right),  \\
\label{singphi6}
n=6, \quad  && w  \cdot   \left( 1-16\,w^2 \right) 
 \left( 1-4\,{w}^{2} \right) 
 \left( 1-{w}^{2} \right)  
  \left( 1-25\,{w}^{2} \right)  \nonumber \\
&&   \left( 1-9\,{w}^{2} \right) 
 \left( 1\, +3w\, +4w^2 \right) \left(1-3w+4w^2\right)  \nonumber \\
&&    \left( 1-10\,{w}^{2}+29\,{w}^{4} \right).
\end{eqnarray}

For $\, n= \,7$ and $\, n=\, 8$, besides modulo primes series calculations described above,
we also generated very large series from which we obtained 
in floating point form, the polynomials given 
in \ref{singphi7phi8} (using generalised differential Pad\'e methods).

If we compare the singularities for $\, \Phi_H^{(n)}$ to those obtained with 
the ``Diagonal model\footnote[8]{Not to be confused with the
``diagonal susceptibility'' and the corresponding~\cite{Diag} $\, n$-fold
integrals $\, \chi^{(n)}_d$.}'' presented 
in~\cite{bo-ha-ma-ze-07}, i.e. $\Phi_D^{(n)}$,
one sees that the singularities of the linear ODE for the ``Diagonal model''
are identical to those of the linear ODE of the  $\, \Phi_H^{(n)}$
 for $n=\, 3, \, 4$ (and are a proper subset to those of
  $\, \Phi_H^{(n)}$ for $n\, =\, 5,\, 6$).
The additional singularities for $n=\, 5,\, 6$ are zeros of the polynomials:
\begin{eqnarray}
n=5, \quad  &&  \left( 1+3\,w+4\,{w}^{2} \right)
 \left( 1+4\,w+8\,{w}^{2} \right)  \nonumber \\
\quad  && \quad  \times 
\left( 1-7\,w+5\,{w}^{2}-4\,{w}^{3} \right), \nonumber \\
n=6, \quad  &&  \left( 1\, +3w\, +4w^2 \right) \left(1-3w+4w^2\right) 
    \left( 1-25\,{w}^{2} \right).  \nonumber 
\end{eqnarray}
For $\, n = 7$, the zeros of the following polynomials (among others)
are singularities which are not 
of Nickel's type (\ref{sols}) and do not occur 
for $\Phi_D^{(n)}$:
\begin{eqnarray}
&& 1 +8w\,  +15w^2\,  -21w^3\,  -60\, w^4 \, 
+16\, w^5\,  +96\, w^6\, +64w^7, \nonumber \\
&& 1 -4w \, -16w^2 \, -48w^3\, 
 +32\, w^4 \, -128\, w^5.  \nonumber
\end{eqnarray}

The linear ODEs of the multiple integrals $\, \Phi_H^{(n)}$ thus
display {\em additional singularities}
 for $n=\, 5,\,  6$ and $n=7$ ($n= \, 8$ 
see below) compared to the
linear ODE of the single integrals $\, \Phi_D^{(n)}$.

We found it remarkable that the linear ODEs
for the integrals $\Phi_D^{(n)}$ display all the Nickelian singularities,
as well as the new quadratic numbers $1+3w+4w^2=\, 0$ found for $\chi^{(3)}$.
It is thus interesting to see how the singularities for $\, \Phi_D^{(n)}$
are included in the singularities for $\Phi_H^{(n)}$ and whether
the new (with respect to  $\Phi_D^{(n)}$) singularities can be given
by one-dimensional integrals similar to  $\Phi_D^{(n)}$.
Let us mention that the  singularities of the linear ODE for $\,\Phi_H^{(3)}$
(respectively $\,\Phi_H^{(4)}$) {\em are remarkably also singularities}
 of the linear ODE for
$\,\Phi_H^{(5)}$ (respectively $\,\Phi_H^{(6)}$).
In the following, we will show how this comes about and how it generalizes.
For this, we solve in the sequel the Landau conditions for
the $n$-fold integrals (\ref{In}).

\subsection{Landau conditions for the $\, \Phi_H^{(n)}$}
\label{subLandaucond}

We remind the reader that the Landau conditions~\cite{bo-ha-ma-ze-07} are
{\em necessary} conditions for singularities to be the
singularities of the {\em integral representation itself}. In a
previous paper~\cite{bo-ha-ma-ze-07}, we have shown for particular
integral representations
belonging to the Ising class integrals~\cite{crandall},
that in fact the solutions of Landau conditions
identify for specific\footnote[5]{In that respect one must recall the notion of 
leading singularities in contrast with the
subleading singularities (see page 54 in~\cite{Smatrix}.)} configurations (see
below) with the singularities of the ODE associated
with the quantity under consideration.

The Landau conditions~\cite{bo-ha-ma-ze-07} amount to carrying out algebraic 
calculations~\cite{bo-ha-ma-ze-07} on the integrand (\ref{In})
to get singularities of these $\, n$-fold integrals or even, as we will see
in the sequel,  singularities of the corresponding linear ODE~\cite{bo-ha-ma-ze-07}.

In the sequel we use the following integral 
representation~\cite{wu-mc-tr-ba-76,nickel-99}:
\begin{eqnarray}
y_j \,x_j^n \,\, =\,\,\, \int_0^{2\pi}\, {d\psi_j \over 2\pi}\cdot 
 {\frac{\exp(i\, n\psi_j)}{1\, -2\, w\cdot (\cos(\phi_j)\, +\cos(\psi_j))}}. 
\end{eqnarray}

Defining
\begin{eqnarray}
D(\phi_j, \psi_j) \, = \,\,\,\,\,
 1\, -2 w \cdot  \left(\cos(\phi_j)+\cos(\psi_j) \right), 
\end{eqnarray}
the integral $\,\Phi_H^{(n)}$ (see its expansion (\ref{integral}) in \ref{a0}),
becomes
\begin{eqnarray}
\label{In-rep}
&&\Phi_H^{(n)} \,=\, \,\,{\frac{1}{n!}}  \cdot
\int_0^{2\pi} \prod_{j=1}^{n} {\frac{d\phi_j}{2\pi}}
{\frac{d\psi_j}{2\pi}}  \\
&& \quad  \quad  \quad  \quad \times \,  D^{-1} \left(\phi_j, \psi_j \right)   \cdot
 \delta \left( \sum_{j=1}^n \phi_j \right) \,
\delta \left( \sum_{j=1}^n \psi_j \right),           \nonumber
\end{eqnarray}
where the Dirac delta's are introduced to take care of the conditions
\begin{eqnarray}
\label{angles}
\sum_{j=1}^n \, \phi_j \,=\,\,0, \qquad \quad 
\sum_{j=1}^n \, \psi_j \,=\,\,0  \quad  \quad
   {\rm mod.} \, \, \,    2 \, \pi
\end{eqnarray}
on both the angles $\,\phi_j$ and the auxillary angles $\, \psi_j$.

The Landau conditions~\cite{Smatrix,itz-zub-80}
 can easily be written~\cite{bo-ha-ma-ze-07}:
\begin{eqnarray}
&& \alpha_j \cdot  D\left( \phi_j, \psi_j \right)\, =\,\, 0, 
\qquad \qquad  \quad j=\,1,\,\cdots, \, n,  \\
\label{pinch}
&& \beta_j \cdot  \phi_j =0, \qquad \quad \gamma_j \cdot  \psi_j\,=\,0, 
\quad \qquad \quad \quad \, j=\,\, 1,\,\cdots, \, n-1,  \\
&& \alpha_j \cdot  \sin(\phi_j) \,
-\alpha_n \cdot  \sin(\phi_n) \, + \beta_j =\,0, 
\qquad \quad j=\,\,1,\,\cdots \, n-1,  \\
&& \alpha_j \cdot  \sin(\psi_j)\,
 -\alpha_n \cdot  \sin(\psi_n)\, + \gamma_j =\,0,
 \qquad \quad j=\,\,1,\,\cdots, \, n-1 
\end{eqnarray}
together with (\ref{angles}).
The Landau singularities are obtained by solving 
these equations\footnote[8]{Note that
conditions (\ref{pinch}), 
$\beta_j \cdot  \phi_j =0, \, \,  \gamma_j \cdot  \psi_j=\,0, 
\, \,  \, j=\,\, 1,\,\cdots, \, n-1$ have to be considered
 in the general Landau conditions. They
do not occur if one restricts oneself to pinch singularities.} in all the
unknowns, where the parameters $\alpha_j$, $\beta_j$, $\gamma_j$
should not be all equal to zero.

In this paper, our aim is not to find all the solutions of the
above equations but to show that the singularities of the linear ODE
for the $\, \Phi_H^{(n)}$ are solutions of the Landau conditions.
Furthermore, in working out various Ising class
 integrals~\cite{crandall}  and the two
models of~\cite{bo-ha-ma-ze-07} (see \ref{bb}), we remarked that the singularities
of the linear ODE are, in fact, included in a particular ``configuration''.
What we mean by ``configuration'' is the set of values (equal to zero or not)
of the parameters $\alpha_j$, $\beta_j$, $\,\gamma_j$.

The ``configuration'' we consider
\begin{eqnarray}
\alpha_j \ne 0, \qquad \qquad \beta_j=\,\, \gamma_j =\,\,0, 
\end{eqnarray}
corresponds to {\em pinch singularities} on the
 manifolds $\,D\left( \phi_j, \psi_j \right)\, = \, 0$. 
One may also be convinced to take $\beta_j =\, \gamma_j =\,0$, since 
the integrand is periodic\footnote[9]{B. Nickel, private communication.}
in $\,\phi_j$ and $\,  \psi_j$.

Let us stress that the the configuration considered
where all the Lagrange multipliers of the 
singularity manifolds $\, D(\phi,\, \psi)$ are different from
zero ($\alpha_j\, \ne \, 0, \,$   for any $\,  j$) leads to the so-called
{\em leading Landau singularities} following the terminology of page
54 of~\cite{Smatrix}.

The Landau conditions become:
\begin{eqnarray}
\label{laneq1}
&& 1\, -2\,w \cdot \left(\cos(\phi_j) +cos(\psi_j) \right) \,=\,\,0,
 \qquad \quad j=\,1,\,\cdots ,\, n,  \\
\label{laneq2}
&& \alpha_j\, \sin(\phi_j)\, -\alpha_n \, \sin(\phi_n) \, =\,\,0,
 \qquad \quad \quad j=\,1,\,\cdots, \, n-1, \\
\label{laneq3}
&& \alpha_j\, \sin(\psi_j)\, -\alpha_n \, \sin(\psi_n) \,=\,\,0, 
 \qquad \quad \quad j=\,1,\, \cdots, \, n-1.
\end{eqnarray}
and:
\begin{eqnarray}
\label{angles2}
\sum_{j=1}^n \, \phi_j \,=\,\,0, \qquad \quad 
\sum_{j=1}^n \, \psi_j \,=\,\,0  \quad  \quad
   {\rm mod.} \, \, \,    2 \, \pi
\end{eqnarray}

The Landau singularities are solutions of these
conditions (see \ref{singLand} for details).
Note that the first three conditions (\ref{laneq1}), (\ref{laneq2}), (\ref{laneq3})
are invariant by the transformation:
\begin{eqnarray}
\label{trans}
w \, \,  \longrightarrow \, \,  -w, 
 \quad \quad \phi_j \, \, \longrightarrow \, \,  \phi_j\, + \pi, 
 \quad \quad \psi_j \, \,  \longrightarrow \, \,  \psi_j\, + \pi.
\end{eqnarray}
but the  Landau conditions (\ref{laneq1}), (\ref{laneq2}), (\ref{laneq3})
together with (\ref{angles2}) are invariant by transformation (\ref{trans}) 
{\em if and only if} $\, n$ {\em is even}. This distinction between 
even and odd integer $\, n$ (corresponding to the symmetry breaking of 
$\, w \, \leftrightarrow \, -w$) is reminiscent of
the distinction between 
even and odd integer $\, n$ for the $\, \chi^{(n)}$ 
associated with the distinction between low and high temperature regimes.

The Landau conditions  yield two families of singularities
expressed  in terms of Chebyshev polynomials of the first
and second kind. The first family reads:
\begin{eqnarray}
\label{famil1corps}
&& T_{2p_1} \left( 1/2w +1 \right)\,  \,=\,\,\,\, \,\, 
 T_{n-2p_1-2p_2} \left( 1/2w-1 \right), \\
&& 0\, \le\, p_1 \,\le \,[n/2], \quad \quad \quad \quad \quad 
 0 \,\,\le\,\, p_2\, \,\le\, \,[n/2]\,\,-p_1 \nonumber
\end{eqnarray}
The second family is given
 by the elimination of $z$ from:
\begin{eqnarray}
\label{famil2corps}
&& T_{n_1}( z)\,\,  
- T_{n_2} \Bigl(  {\frac{4w-z}{1-4w\,z }} \Bigr) \,\, = \,\,\,  0,  \\
&& T_{n_1} \left( {1 \over 2w}- z \right)\,\, \,
 - T_{n_2} \Bigl( {1 \over 2w}- {\frac{4w-z}{1-4w\,z }} \Bigr) 
\,\, = \,\, \, 0, \nonumber \\
&& U_{n_2-1} (z) \cdot  
U_{n_1-1}\Bigl( {1 \over 2w}- {\frac{4w-z}{1-4w\,z }} \Bigr)\nonumber \\
&& \qquad  \qquad 
- U_{n_2-1}\left( {1 \over 2w}- z \right)\cdot 
U_{n_1-1} \Bigl(  {\frac{4w-z}{1-4w\,z }} \Bigr) \,\,=\,\,\, 0  \nonumber 
\end{eqnarray}
with
\begin{eqnarray}
\label{part-integers}
&&  n_1 \,=\, p_1, \qquad \qquad  n_2 \,=\,\, n\,-p_1\,-2p_2,   \\
&&  0 \,\,\le \,\,p_1\, \le\, n, \quad \quad \qquad 
0\,\, \le\,\, p_2 \,\, \le\, \,[(n-p_1)/2].
\end{eqnarray}

One recognizes in the first set of equations (\ref{famil1corps}),
a generalization of the singularities given by Nickel \cite{nickel-05}
for the pinch singularities coming from the product of the $y_j$'s,
and also derived for our multiple integral 
denoted $\, Y^{(n)}$ in~\cite{bo-ha-ma-ze-07}. These have been written
as~\cite{bo-ha-ma-ze-07,nickel-05}:
\begin{eqnarray}
\label{Yn}
 T_{k} \left( 1/2w+1 \right)\,\,  \,=\,\,\,\, \,\, \,
 T_{n-k} \left( 1/2w-1 \right)
\end{eqnarray}
Note that, comparatively to (\ref{famil1corps}),  the integer $k\,$
should be even\footnote[1]{This is a 
consequence of (\ref{laneq1}), (\ref{laneq2}), (\ref{laneq3}), (\ref{angles2})
yielding $\, k \cdot \pi \, = \, 0$ mod. $\, 2 \, \pi$ (see \ref{D1}).}. 

The second set of equations (\ref{famil2corps}) is
a generalization of the singularities we derived for $\, \Phi_D^{(n)}$
in~\cite{bo-ha-ma-ze-07}.
In both formulae, one notes the occurrence of a second varying integer $\, p_2$,
leading to a better understanding of the singularities of these
integrals.
Indeed with $\, p_2$ running, the  linear ODE for $\, \Phi_H^{(n)}$  will
automatically  contain
all the singularities of the  linear ODEs for
 $\, \Phi_H^{(n-2)}$, $\, \Phi_H^{(n-4)}$, $\, \cdots $, 
$\, \Phi_H^{(n-2q)}$.

For $\, n=\, 7$, we have checked that the singularities
specific to $\, n=\, 7$ ($p_2=0$ in (\ref{famil1corps}),
 (\ref{famil2corps})) also appear as
singularities of the linear ODE in floating point form (see
Appendix D for details). For $\, p_2=\, 1$, part of the 
singularities appear in floating point form, while for
$\, p_2=\, 2$ (i.e. singularities of $\, \Phi_H^{(3)}$), no 
singularities appear in floating point form.
Similarly, for $\, n=\, 8$, we have checked that the
singularities specific to $\, n=\, 8$ ($p_2=\, 0$
 in  (\ref{famil1corps}), (\ref{famil2corps}))
also appear as singularities of the linear ODE in floating point form (see
Appendix D for details). For $p_2\, \ge \, 1$,  no 
singularities appear in floating point form.

Let us remark that the non observation of some
singularities in floating point form is not really
significant. Indeed, we have used 1250 (resp. 1200
terms) for  $\, \Phi_H^{(7)}$ (resp. $\, \Phi_H^{(8)}$) while the 
$\, \Phi_H^{(7)}$ and $\, \Phi_H^{(8)}$ linear ODEs 
need more than 20000 terms.

\begin{figure}
\psfig{file=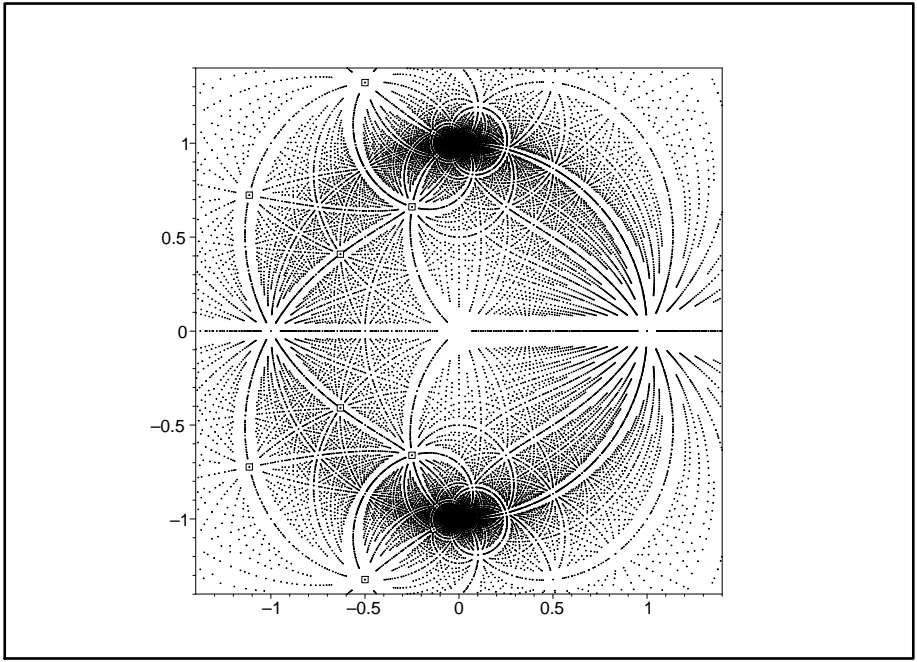,scale=0.53,angle=-90}
\caption{First family of singularities (\ref{famil1corps})
 in the complex $\, s$ plane ($n \, \le \, 51$).}
\label{f:fig1}
\end{figure}

\begin{figure}
\psfig{file=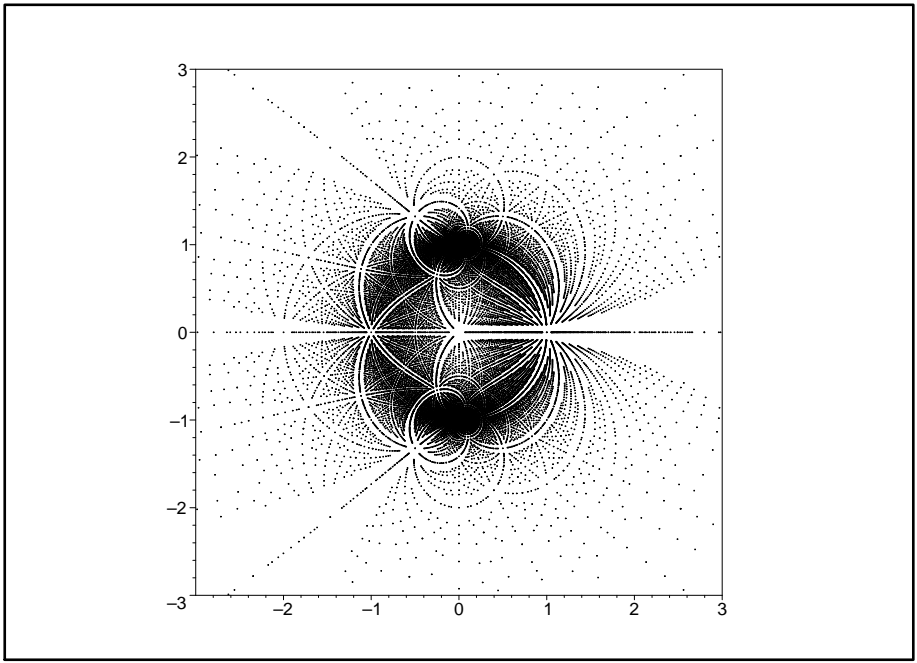,scale=0.53,angle=-90}
\caption{First family of singularities (\ref{famil1corps}) in the complex $\, s$ plane
 far from the unit circle ($n \, \le \, 51$).}
\label{f:fig2}
\end{figure}

\begin{figure}
\psfig{file=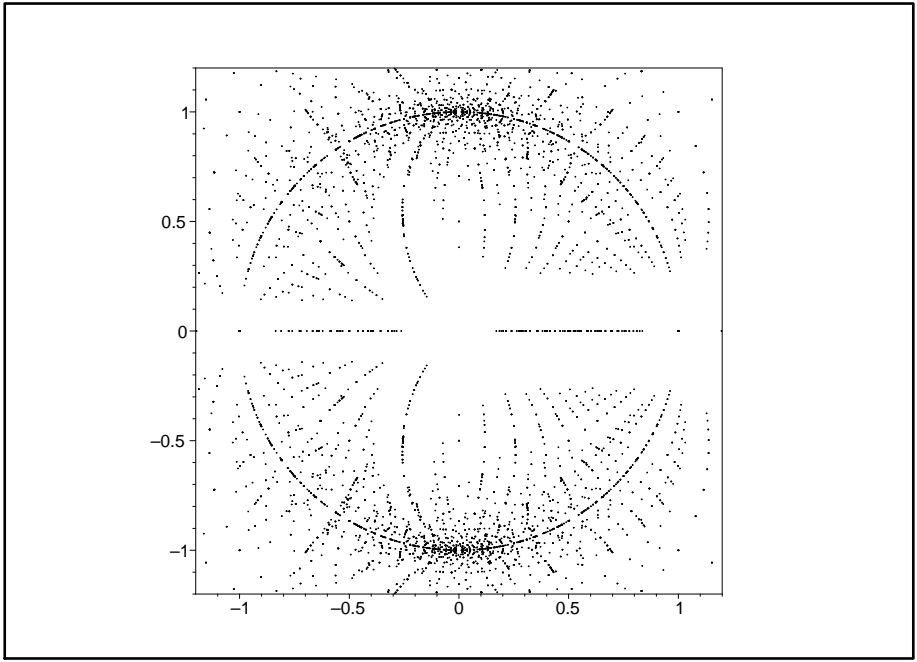,scale=0.53,angle=-90}
\caption{First and second family of singularities (\ref{famil1corps}),
 (\ref{famil2corps})  in the complex $\, s$ plane ($n \, \le \, 16$).}
\label{f:fig3}
\end{figure}

\begin{figure}
\psfig{file=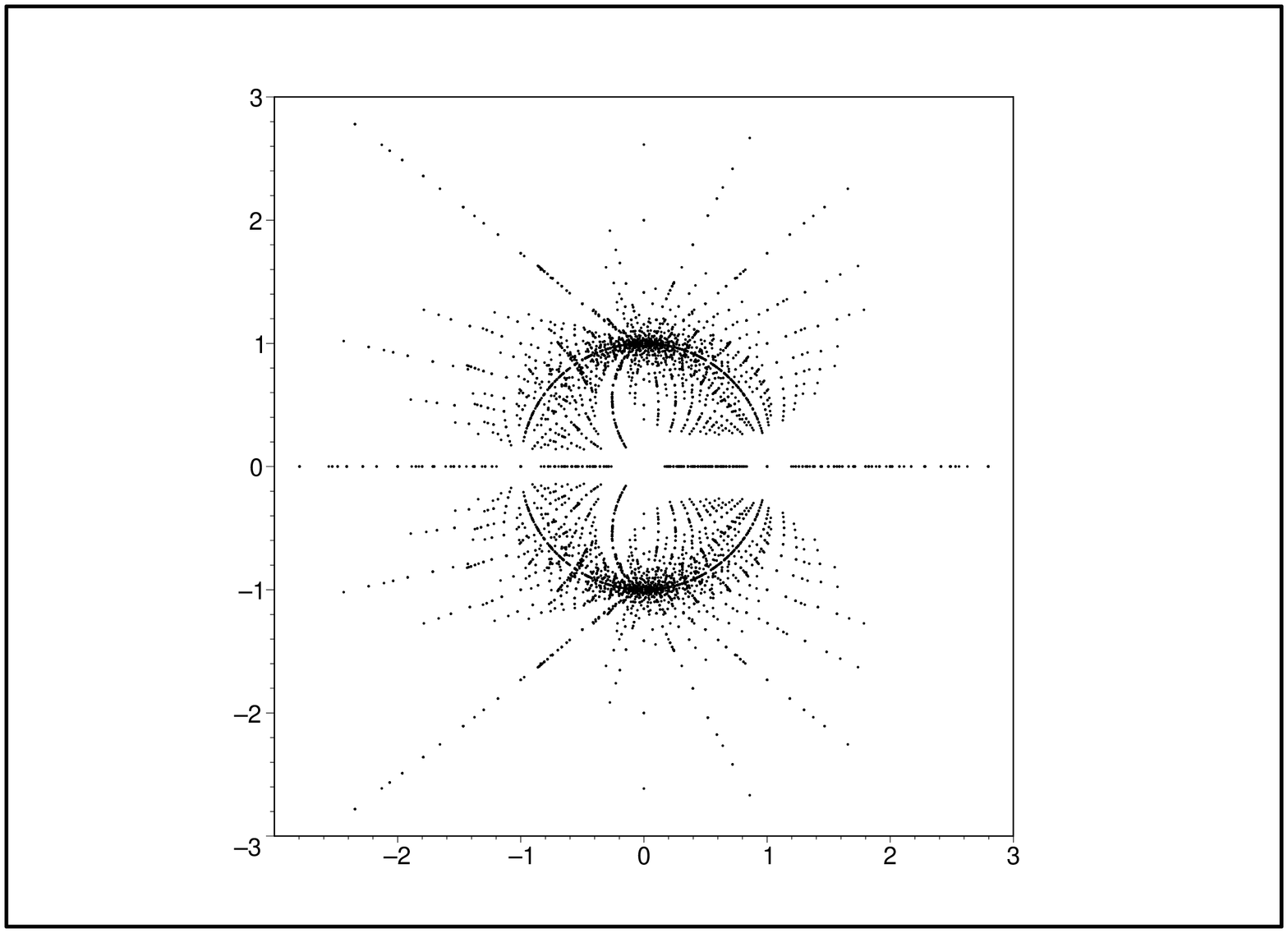,scale=0.53,angle=-90}
\caption{First and second family of singularities (\ref{famil1corps}), 
 (\ref{famil2corps}) in the complex $\, s$ plane far from the unit circle ($n \, \le \, 16$).}
\label{f:fig4}
\end{figure}

\vskip .2cm 
Figure \ref{f:fig1} shows the first family of singularities (\ref{famil1corps}) 
displayed in the complex $\, s$ plane 
close enough to the unit $\, s$-circle. This figure clearly 
shows a quite rich structure for these set of points. 
This figure looks like a network of nodal points linked together by 
(cardioid-like) curves
 that can, at first sight, hardly be distinguished from arcs of circles. 
In particular the selected points $\, 1+3\, w + 4 w^2 \, = \, 0$ 
as well as the singularities for $\, \Phi^{(5)}$, like 
$\, 1+8\, w \, + 20 w^2 +15\, w^3 \,+4 \, w^4  \, = \, 0$ 
can be seen to occur quite clearly as some of these nodal points.

Figure \ref{f:fig2}  shows the first family of singularities (\ref{famil1corps}) 
far  from the unit circle. 
Figure \ref{f:fig3}  shows all the singularities altogether (first and second family)
close to the  unit $\, s$-circle.
Finally figure \ref{f:fig4}  shows all the singularities
 together ((\ref{famil1corps}), (\ref{famil2corps}))
that are not so close to the  unit $\, s$-circle.
\vskip .2cm 

The accumulation of singularities one can see
 on figure \ref{f:fig1}  near $\, s\,=\,i $ 
and $\, s\,= \,-i $  seem to confirm 
the statement made in Orrick {\it et al}~\cite{Orrick} that 
these two points are two quite unpleasant points for the susceptibility
of the Ising model for which the series expansions 
are not even asymptotically convergent.

Besides reproducing exactly the singularities of the linear ODE for
$\, \Phi_H^{(n)}$, it is remarkable to see from the formula
(\ref{famil1corps}), (\ref{famil2corps}),
how to track where each singularity-polynomial comes from.
This allows one to understand how the singularities of the Ising like
integrals $\, Y^{(n)}$ and $\, \Phi_D^{(n)}$
(see \cite{bo-ha-ma-ze-07}) and even the Nickelian
singularities (\ref{sols}) emerge in these multiple integrals (\ref{In}).
This comes simply from the partition (\ref{part-integers}) and the
equivalent one in (\ref{famil1corps}).

\subsection{Singularities: from $\, n$-fold integrals 
to one dimensional integrals}
\label{land}

Consider for instance the singularities $\, 1-7w+5w^2-4w^3\,=\,0$ occurring
in $\, \Phi_H^{(5)}$, which are given by (\ref{famil1corps}) for
$n=\,5$,  $p_1=\, 1$ and $p_2=\, 0$.
As far as conditions
on the integration angles (see (\ref{phi-diag}) below), this arises
 from a situation where two angles are equal and
the three others are equal. 
Recall that the $\, \Phi_D^{(n)}$ integrals are constructed with
the following restrictions on the angles:
\begin{eqnarray}
\label{phi-diag}
\phi_1\,=\,\, \phi_2\, =\, \,\cdots \,\,=\,\phi, \qquad \quad
 \phi_n\,=\,\, -(n-1)\, \phi.
\end{eqnarray}
One sees that a generalization of this model (\ref{phi-diag}) is simply:
\begin{eqnarray}
\label{diag-gener1}
\phi_1\,=\,\,\phi_2\, = \,\,\cdots\,\, =\,\,\phi_k,
\nonumber \\
\phi_{k+1}\,=\,\,\phi_{k+2}\, =\, \,\cdots\,\, =\,\,\phi_n, 
\qquad \quad 
k =\,\,\, 0,\, 1, \,\,\, \cdots,\,\, \,[n/2].
\end{eqnarray}
By the condition on the angles, {\em this case is 
indeed one dimensional}, with:
\begin{eqnarray}
\label{diag-gener2}
\phi_n \,=\,\, -{(n-k) \over k} \cdot \phi\,  \,\,  +{2 j\,\pi \over k}, 
\,\quad \quad \quad \quad 
j\,\,\,\,\, {\rm integer}.
\end{eqnarray}

The model (\ref{phi-diag}) is obviously given by
  (\ref{diag-gener1}) for
$k=1$. The Nickelian singularities are also given 
 by (\ref{diag-gener1}) for
$k=0$, but this time, the underlying model is ``zero-dimensional''.
The model constructed along the same lines as 
in~\cite{bo-ha-ma-ze-07} corresponds to an integrand:
\begin{eqnarray}
\label{moresimple}
\sum_{j=0}^{n-1} \, \, {{1} \over {
 1 \, - x^n \left( {2\pi\,j \over n} \right) }}.
\end{eqnarray}
The Nickelian singularities arise as poles.

For $\, k \ge 2$, the singularities given by the
 model (\ref{diag-gener1}), which
appear in (\ref{In}), are thus given neither by (\ref{sols}) nor by
$\Phi_D^{(n)}$.
Consider one variable of integration such as (\ref{chinaked}), where 
the integrand is:
\begin{eqnarray}
\label{diag-gener}
{{1} \over { 1\,\, -x^{n-1}(\phi)  \cdot  x ((n-1)\phi) }}
\,\,\quad 
\longrightarrow  \quad 
\,\,
{{1} \over {  1\,\, -x^{n-k}(\phi)  \cdot  x^k (  \phi_n )  }}.
\end{eqnarray}
and denote by $\,\Phi_k^{(n)}$ such integrals
(one then has  $\,\Phi_1^{(n)} =\,\Phi_D^{(n)}$).

Fix $\, n=\, 5$ and $k=\, 2$.  The constraint 
(\ref{diag-gener2}) on the angles reads:
\begin{eqnarray}
\label{phi5}
\phi_5\, =\,\, -{3 \over 2}\, \phi_1 \,\, +\, j\,\pi, 
\quad \quad \quad 
j\,\, {\rm integer}
\end{eqnarray}
with one integration variable.
The series of coefficients of  $\,\Phi_2^{(5)}$ is generated along 
 the same lines as for $\Phi_D^{(n)}$
(see \ref{a0}). The Fuchsian linear 
differential equation is of order six
and this order is independent of the value of $j$ in (\ref{phi5}).
The singularities of the linear ODE are zeros of the following polynomials:
\begin{eqnarray}
&& w  \cdot   \left( 1-16\,w^2 \right)  \left( 1+w \right)
 \left( 1-3\,w+{w}^{2} \right)  \left( 1+2\,w-4\,{w}^{2} \right) \nonumber \\
&& \quad \quad \quad \times  \left( 1+4\,w+8\,{w}^{2} \right)\,
  \left( 1-7\,w+5\,{w}^{2}-4\,{w}^{3} \right).
\end{eqnarray}
We obtain singularities (from the last two polynomials) appearing for
$\Phi_H^{(5)}$ and not occurring for $\Phi_D^{(5)}$.

The occurrence of the singularities $ \,1+3w+4w^2 \,= \,0 $ for 
(the linear ODE of) $\,\Phi_H^{(5)}$
{\em but not for} (the linear ODE of) $\,\Phi_D^{(5)}$ 
is explained along similar lines.
Note that these singularities are common to (the linear ODE of) $\,\Phi_H^{(3)}$,
$\,\Phi_H^{(5)}$ and $\,\Phi_H^{(6)}$.
The polynomial $ \,1+3w+4w^2 \, $ appears
 for (the linear ODE of) $\,\Phi_H^{(5)}$ from
(\ref{famil1corps}), namely:
\begin{eqnarray}
 T_{2p_1} \left( 1/2w+1 \right) \,\,=\,\,\,\,
\, T_{n\,-2p_2\,-2p_1} \left( 1/2w-1 \right).
\end{eqnarray}
One sees that the polynomial $ \,1+3w+4w^2 \,$ will appear for all
combinations of $n$, $p_1$ and $p_2$  such that:
\begin{eqnarray}
\label{fulfill}
 2\,p_1 \, =\,\, 2, \qquad \quad  n\,-2\,p_2 \,-2\,p_1\, \,=\,\, \,1.
\end{eqnarray}
In other words, the polynomial that arises for given $n$ and $p_1$, {\em will
also appear} for the same value of $p_1$ and for $n-2p_2$.
The singularities corresponding to $ \,1+3w+4w^2 \,=0 $ occur for $\Phi_H^{(5)}$
with $n=\, 5$, $p_1=1$ and $p_2=1$, but  (\ref{fulfill}) is
also satisfied for  $n=3$, $p_1=1$ and $p_2=0$ which shows a situation
with three angles,with two of them equal. This is precisely the integrand
in (\ref{chinaked}), i.e. in $\Phi_D^{(3)}$.

Consider now the case $n=\, 6$ and $k=\, 2$. This amounts to
considering the $\,n$-fold integral $\, \Phi_2^{(6)}$ 
with: 
\begin{eqnarray}
\label{phi6}
\phi_6 =\, \,  -2\, \phi_1  \,\,+\, j\,\pi,
 \quad \quad \quad \quad
 j\,\, {\rm integer}.
\end{eqnarray}
The results are dependent on the integer $j$. For instance, the series
around $w=0$ reads:
\begin{eqnarray}
\label{serplusminus}
&&\Phi_2^{(6)}\, =\,\, 1+w^6 + 32 w^8 \pm w^9 \, + 659\, w^{10} \,  \\
 && \quad \quad \quad \qquad \qquad \pm 1296 w^{11}\,
 +11691\, w^{12}\, + \, \cdots \nonumber
\end{eqnarray}
With the $+$ sign in the series
(\ref{serplusminus}), the linear differential
 equation is of order five and the
singularities are given by the zeros of the polynomials:
\begin{eqnarray}
 && w  \cdot   \left( 1-16\,w^2 \right)
 \left( 1-w \right)  \left( 1+2 w \right)  
  \left( 1-9\,{w}^{2} \right) \nonumber \\
&& \qquad  \qquad \quad \times   \left( 1-25\,{w}^{2} \right)\,
  \left( 1\, +3w\, +4w^2 \right).
\end{eqnarray}
The results corresponding to the choice of a minus sign in the series
(\ref{serplusminus}) are obviously obtained
by\footnote[3]{The last case for $n=6$, i.e. $k=3$
 does not provide singularities
other than Nickel's.} $\,w \, \rightarrow \, -w\,$.
We obtain the singularities $ 1-25\,{w}^{2}$ and $\, 1 \pm 3w +4w^2=0$
occurring for (the linear ODE of) $\,\Phi_H^{(6)}$ 
but not for (the linear ODE of) $\, \Phi_D^{(6)}$. 

Similarly, for $n=\, 7$, ($k$ goes to 3), one 
obtains for $k=2$, the singularities
as zeros of the following polynomial
$\,1+8w+15w^2-21w^3-60w^4+16w^5+96w^6+64w^7$, which has indeed been
found numerically in the linear ODE search on a 
large series corresponding to $\,\Phi_H^{(7)}$ (see \ref{singphi7phi8}).

We have the remarkable fact that the singularities of the linear ODE for
the multiple integral $\, \Phi_H^{(n)}$ are 
given by a finite set of singularities 
of linear ODEs of a set of one-dimensional integrals,
 namely, $N(N+1)/2$ one-dimensional integrals,
with $N=\, [n/2]$. For instance, the singularities of the four-dimensional
integral $\,\Phi_H^{(5)}$ identify with those of, at most, three one-dimensional
integrals. This appears, simply, from the couple of integers in
(\ref{famil1corps}) which read $\left(2p_1, \, n-2p_2 \, -2p_1 \right)$.
For fixed $\,n$, when $\, p_2$ varies, one sees that we are in fact considering 
all the lower integer values $\,n-2p_2$. The same situation holds for (\ref{famil2corps}).
This identification leads, obviously, to particular structures in the
singularities for different $ \, n$. This is what we show in the sequel.

\subsection{Singularity structures of $\, n$-fold integrals 
and particular sets of one-dimensional integrals}
\label{singversus}

The Landau singularities given in \ref{singLand} are checked against the
singularities of the linear ODE for $\, \Phi_H^{(n)}$
($n\,=\,3,\,\cdots,\, 6$), and {\em are found to be identical}.
Assume that these formulae do indeed reproduce all the singularities
of the linear ODE for $\Phi_H^{(n)}$, for any $\, n$.
In this case, we can check whether the singularities appearing at $n=m$ 
also occur for $\, n=\, \,m+1,\, \,  n=\,\, m+2, \, \,\cdots $

We have found that the singularities at  order $2\,n$ will {\em also} be
singularities at order $2\,n + 2\,p$, where $p$ is a positive integer.
Similarly, the singularities at order $\, 2\,n + 1$ will also be present
at the following odd orders.

What is remarkable is the fact that the singularities at odd order also appear
at even orders. The rule is: {\em all the singularities at odd order
$n$ also appear in the higher orders (odd and even) except for the first
$(n-1)/2$ even orders}. For instance, the singularities appearing at $n=3$
will occur for all $n$, except the first even order, i.e. $4$.
The singularities appearing at $n=5$ will occur for all $\, n$, except the
first two even orders, i.e. $6$ and $8$.

The consequence of this {\em  embedding} of the singularities is the
occurrence of some singularities at {\em predefined orders}.
The singularity $1+2w=0$ is present {\em at any order} $n$.
The singularity $1-2w=0$ is present for any even order $2\,n$.
The singularity $1+w=0$  occurs at any order $n \ge 5$.
The singularity $1-w=0$  occurs at any order $n$, except for $n=4$.
All these singularities are Nickelian.
The first non Nickelian singularity $\, 1\, +3w\, +4w^2 \, =\, 0$ {\em appears
at all orders} $\, n$, except for $n=4$.

Moreover, we have given in~\cite{bo-ha-ma-ze-07}
the Landau singularities for the (linear ODEs of the) integrals $\Phi^{(n)}_D$.
These singularities have been found to be identical with the singularities
of the linear ODE for $\Phi^{(n)}_D$ obtained exactly up to $n=8$ and
modulo a prime up to $n=\,14$.
We have seen that all the singularities of the linear ODE
of $\Phi^{(n)}_D$ in the variable $s$ lie in the annulus defined
by two concentric circles of radius $\sqrt{2}$ and $1/\sqrt{2}$.
The radii of the two concentric circles are the roots,
in the variable $s$, of the polynomial $1+3w+4w^2=\, 0$, 
that is $\, s^2+s+2\, = \, 0$
and $\, 1+s+2\,s^2 = \, 0$. 
With the multiple integrals $\,\Phi_H^{(n)}$, one sees that some of the
singularities {\em are not confined} to this annulus anymore.

Thanks to the Landau conditions, one can now understand this structure
from the reduction of the multiple integrals $\, \Phi_H^{(n)}$
to a set of one-dimensional integrals $\, \Phi_k^{(n)}$ as far as the
location of singularities is concerned.
For $k=\, 0$, which corresponds to the Nickelian singularities, the ``annulus''
is the unit circle. For $k=1$ corresponding to the integrals
$\, \Phi_D^{(n)}$, one has the annulus of radii $\sqrt{2}$ and its inverse.
For each $k$, one expects the singularities to lie in 
an annulus with a concentric structure. For these
annuli the larger radius increases (smaller radius decreases) as $k$
increases.
From the reduction of the singularities of
$\, \Phi_H^{(n)}$ to  these $\, \Phi_k^{(n)}$,  all the singularities for fixed
$p_1=\, k$ in (\ref{famil1corps}) and for fixed $p_1=k$ in (\ref{famil2corps})
will be confined to one annulus.
For instance for $k=\, 2$, all the singularities occurring in the linear ODE
for $\, \Phi_k^{(n)}$, (i.e. for all $n$), or, equivalently,  all the
singularities obtained by (\ref{famil1corps}) for $p_1=\, 1$ and
by (\ref{famil2corps}) for $p_1=\, 2$ will be confined to the annulus
of radii $2.79\, \cdots $ and its inverse. This value is the root, in the
variable $s$, of $\,  1-7\,w+5\,{w}^{2}-4\,{w}^{3}  \,=\,0$
occurring for $\, \Phi_H^{(5)}$. For $k=\, 3$, one remarks that the
annulus will not be obtained from (\ref{famil1corps}) which is
restricted by $2p_1$, an even integer. In fact this is general.
The radii of the annuli are given by (\ref{famil1corps}) for $k$ even
and by (\ref{famil2corps}) for $k$ odd. The root in the variable $s$
that will define the annulus occurs at odd order $n$ given by $2k+1$.

The picture now, is as follows. The singularities of the linear ODE for
the integrals $\, \Phi_H^{(n)}$ are partionned into ``families'' indexed
by the integer $k$. The singularities for $k=0$ are Nickelian and lie
on the unit circle, say, $r_0=1$. The singularities for $k=1$ lie in the
annulus $r_1=\, \sqrt{2}$, $\, 1/\sqrt{2}$ (we discard from now on, the
smaller radius). The singularities for $k=\, 2$ will be confined in the
annulus $r_2$. The singularities for $k=N$ will be in the annulus $r_N$.
These concentric annuli are such that $r_0 < r_2 < \cdots < r_{2N}$
and $r_1 < r_3 < \cdots < r_{2N+1}$, (with $r_{2k} < r_{2k+1}$).
As $k$ grows, the radii of two neighboring circles behave as
$r_{2k+2}-r_{2k} \rightarrow 0$ and $r_{2k+3}-r_{2k+1} \rightarrow 0$.
This decrease is not enough to create
an accumulation of circles. We checked with $k=300$ circles that the decrease
goes as $k^{-\alpha}$ with $\alpha <1$ preventing any convergence.
For $\, n$ large these radii diverge: $\, r_{N}\, \, \rightarrow \, \, \infty$ 
when $\, N\,\, \, \rightarrow \, \, \infty$. 

Note that these families, (i.e. the index $k$) come from the resolution
of the Landau conditions and from the reduction of the singularities
for  $\, \Phi_H^{(n)}$ to the ones of $\, \Phi_k^{(n)}$, ($k=0,1, \cdots [n/2]$).
We have no idea as to how these families can be seen directly from
the mutiple integrals $\, \Phi_H^{(n)}$.
If the singularities for $\, \Phi_H^{(n)}$ happen to
be identical with those occurring in the linear ODE for $\, \chi^{(n)}$,
it may become important to see whether this picture persists
and whether this picture is showing another partition of the
susceptibility $\, \chi$ instead of the known sum on $\, \chi^{(n)}$.

\begin{figure}
\psfig{file=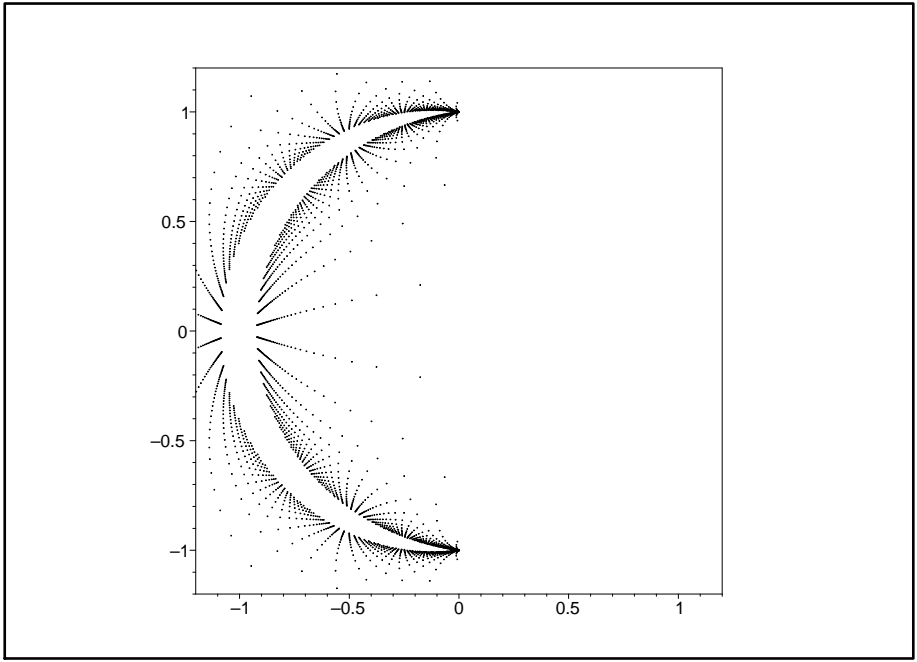,scale=0.53,angle=-90}
\caption{Crescent in the complex $\, s$ 
plane given by
 (\ref{famil1corps}): $\, k\, = \, 5$, $\, n \, \le \, 91$, $\, n$ odd.}
\label{f:fig5}
\end{figure}

\begin{figure}
\psfig{file=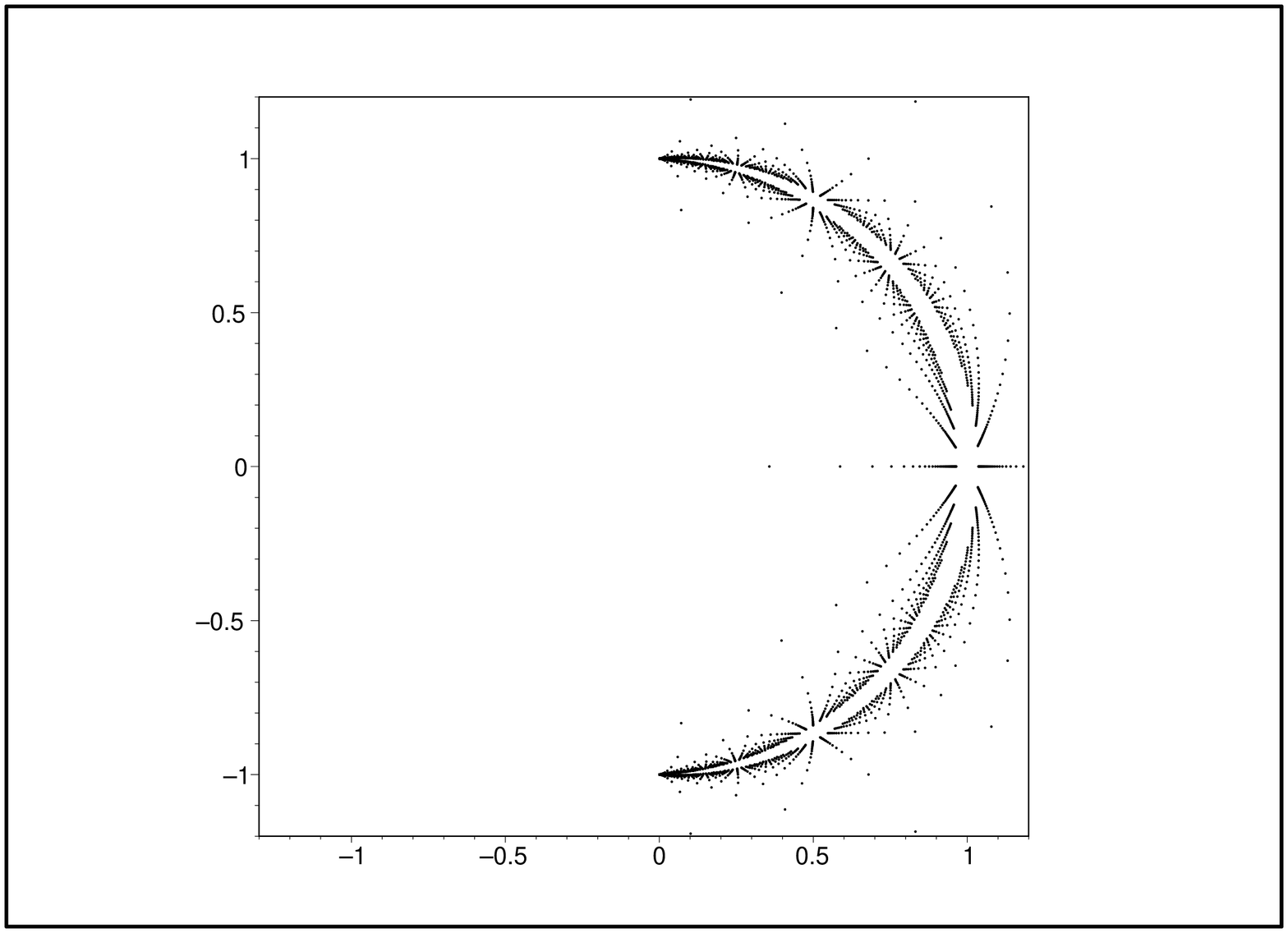,scale=0.53,angle=-90}
\caption{Crescent in the complex $\, s$
 plane given by 
(\ref{famil1corps}): $\, k\, = \, 2$, $\, n \, \le \, 71$, $\, n$ odd.}
\label{f:fig6}
\end{figure}

\begin{figure}
\psfig{file=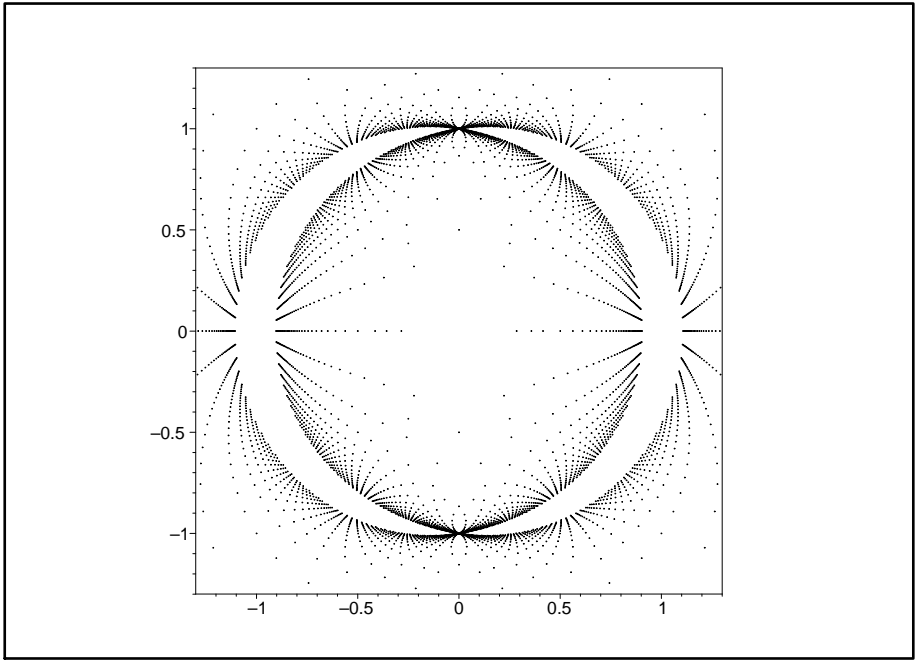,scale=0.53,angle=-90}
\caption{Crescent in the complex $\, s$ 
plane given by 
(\ref{famil1corps}): $\, k\, = \, 6$, $\, n \, \le \, 80$, $\, n$ even.}
\label{f:fig7}
\end{figure}

\vskip .2cm 
 Figure \ref{f:fig5}, \ref{f:fig6}  and \ref{f:fig7} 
 show how the first family of singularities (\ref{famil1corps}) 
 in the $\, s$ complex plane is decomposed according to the integer $\, k$ 
in (\ref{diag-gener1}).  Figure 5 shows singularities (\ref{famil1corps}) 
for a {\em given odd value} of $\, k$, namely $\, k\, = \, 5$ {\em for 
any odd values} of $\, n$ up to $\, 91$.  Figure \ref{f:fig6}
 shows singularities (\ref{famil1corps}) 
for a given even value of $\, k$, namely $\, k\, = \, 2$ for 
any odd values of $\, n$ up to $\, 71$. Figure 7 
shows singularities (\ref{famil1corps}) 
for a given even value of $\, k$, namely $\, k\, = \, 6$ for 
any even values of $\, n$ up to $\, 80$. 
The figures corresponding to the filtration of the singularities
 of the first family (\ref{famil1corps}) 
in terms of the integer $\, k$ (previously displayed 
altogether with figures \ref{f:fig1} and \ref{f:fig2})
deserve some comments. First, one sees that the various ``crescent'' 
corresponding to different values of $\, k$ are very similar. 
Secondly one sees from figure 5 that the odd $\, n$, odd $\, k$ ``crescent'' 
break the $\, s\, \leftrightarrow \, -s$
  symmetry (for even $\, n$, even $\, k$,
the equations for the set of singularities
 are functions of $\, s^2$, see figure \ref{f:fig7}) in a quite 
dramatic way: the singularities in
the ``crescent''  of figure \ref{f:fig5} all lie
 {\em only in the left half} $\, s$-complex plane. 
Similarly the singularities in the 
``crescent''  of figure \ref{f:fig6} all lie in the right half  $\, s$-complex plane. 

Along this $\, s\, \leftrightarrow \, -s$  symmetry line 
it is worth recalling that the low-temperature susceptibility
of the Ising model has this $\, s\, \leftrightarrow \, -s$  symmetry
(the low-temperature susceptibility is a function of $\, s^2$ or $\, w^2$)
but the high-temperature susceptibility breaks that 
$\, s\, \leftrightarrow \, -s$  symmetry, and this is also the case
for the $\, n$-fold integral $\, \chi^{(n)}$ with $\, n$ odd.
Our $\, n$-fold integrals (\ref{In}) are introduced to provide an educated guess 
as to the location of the singularities of the  $\, \chi^{(n)}$.
As far as location of singularities of the  $\, \chi^{(n)}$ are concerned,
 it is not totally clear for $\, n$ odd if the $\, s\, \leftrightarrow \, -s$ 
(resp. $\, w\, \leftrightarrow \, -w$)  symmetry 
will not be partially restored on the global set of singularities
with the occurence for a singularity $\, P_n(w) \, = \, 0$ 
for a given value of $\, n$, of the opposite value for, perhaps, a different value
of $\, n$ : $\, P_m(-w) \, = \, 0$. 

\vskip .1cm 
{\bf Remark:} Quite often, in this paper, we use (by abuse of language) the words 
``singularities of an $\, n$-fold integral'' to describe a larger set of singularities,
 namely the singularities of the linear ODEs that the  $\, n$-fold integral 
satisfies. A rigorous study would require, for any ``singularity'', to perform the 
(differential Galois group and connection matrix) analysis
 we have performed in~\cite{ze-bo-ha-ma-05c}. It amounts to getting extremely large series, 
deduced from the obtained linear ODE, that coincide with the series 
expansion of the  $\, n$-fold integral we are interested in, and find out if these
series actually exhibit these singularities. With this tedious, but straightforward,
procedure we can extract the singularities of a specific $\, n$-fold integral
among the larger set of singularities of the corresponding linear differential equation. 
In view of the large number of singularities we display in this
paper, we have not performed such a systematic analysis, that would have been 
quite huge. Furthermore it is important to note that this ``connection matrix''
approach~\cite{ze-bo-ha-ma-05c} requires to have the linear ODE of the $\, n$-fold integral. 
A knowledge of the  linear ODE modulo a prime {\em is not sufficient}. 
We could have performed this analysis for $\, \Phi^{(3)}_H$
and $\, \Phi^{(4)}_H$, but, in that case, we already have 
a deeper result~\cite{ze-bo-ha-ma-05c} namely
the  connection matrix analysis for $\, \chi^{(3)}$
and $\, \chi^{(4)}$, providing an understanding of the singularities
of these $\, n$-fold integrals themselves (in $\, w$ and {\em also} in $\, s$).  

Right now, the only singularities found for the $\, \chi^{(n)}$,
other than Nickelian,  are the quadratic roots of $1+3w+4w^2=\, 0$,
(i.e. the first annulus) which appear at all orders (except $\, n=4$
for $\, \Phi_H^{(n)}$).
Let us show, in the sequel, how this polynomial can be ``special''.

\section{Towards a mathematical interpretation of the singularities}
\label{bridge}

In a set of papers~\cite{Painleve,Fuchs}, we have underlined
the central role played by the {\em elliptic parametrization}
of the Ising model,
in particular  the role played by the second order linear
differential operators  corresponding  to the complete elliptic
integral $\, E$ (or  $\, K$), and the occurrence of an infinite 
number of  {\em modular curves}~\cite{Holo},
canonically associated with {\em elliptic curves}.
The deep link between the theory of elliptic curves and the theory
of modular forms is now well established~\cite{rhoades}.

Consequently, it may be interesting to seek  ``special values'' of
the  modulus $\, k$, (singularities of the $\chi^{(n)}$)
that might have a ``physical meaning'',
as well as a ``mathematical interpretation''.

For that purpose,  recall that 
the modular group requires one to introduce the elliptic nome, defined in terms 
of the periods of the elliptic functions, 
\begin{eqnarray}
\label{defq}
q \, =\,\,\, \exp \Bigl(-\pi \, {\frac{K(1-k^2)}{K(k^2)}}  \Bigr)
 \,=\,\,\, \exp(i\, \pi \, \tau)
\end{eqnarray}
and the half period ratio\footnote[5]{In the theory of
modular forms $q^2$ is also sometimes used instead of $\,q$.
In number theory literature the half-period ratio is
taken as $-i\,\tau$.} $\,\tau$. We write the complete elliptic integral
$K$ as
\begin{eqnarray}
\label{Kk}
K(k) \,=\, \,\,\,\,
 {_2}F_1 \Bigl( 1/2, 1/2; 1; k \Bigr).
\end{eqnarray}
Relations between $\, K(k)$ evaluated at two different moduli
can be found in, e.g. ~\cite{erdeleyi}.

\subsection{Some isogenies of elliptic curves seen as 
generators of the renormalization group}
\label{isogenies}

 The arguments in $\, K$ in these
identities are related by  the so-called, respectively, descending Landen
and ascending Landen (or Gauss) transformations:
\begin{eqnarray}
\label{landenD}
 k \, \quad \longrightarrow \,\quad  
 k_{-1}\, = \,\,\, \, {\frac{1-\sqrt{1-k^2}}{1+\sqrt{1-k^2}}}
\end{eqnarray}
\begin{eqnarray}
\label{landenA}
 k \,\quad  \longrightarrow \,\quad 
 k_{1}\, =\, \,\,\,  {{ 2 \sqrt{k}} \over {1+k}}
\end{eqnarray}
These transformations (or correspondences~\cite{Veselov,Veselov2}), 
decrease or increase the modulus respectively.
Iterating (\ref{landenD}) or (\ref{landenA}), one
converges to
$\, k\, = \, 0$ or $\, k\, = \, 1$ respectively. The half period ratio 
transforms through (\ref{landenD}), (\ref{landenA}),
 as
\begin{eqnarray}
  \tau  \, \rightarrow \, 2\, \tau,  \qquad \quad \quad 
  \tau  \, \rightarrow \, {\frac{1}{2}}\, \tau
\end{eqnarray}
respectively.
The {\em real} fixed points of the transformations
 (\ref{landenD}) and (\ref{landenA})
are $\, k\, = \, 0$
 (the trivial infinite or zero temperature points) and
$\, k\, = \, 1$ (the ferromagnetic and antiferromagnetic
critical point of the square Ising model). 
In terms of the half period ratio,
this reads $\tau=\infty$ and $\tau=0$ respectively,
which {\em correspond to a degeneration
 of the elliptic parametrization
into a rational parametrization}.
In view of these fixed points,  it is natural to identify the
transformations (\ref{landenD}) or (\ref{landenA}), and more generally
any transformation\footnote[3]{See relation (1.3) in~\cite{Schwartz}.} 
$\, \tau \, \rightarrow \, n \cdot \tau$
or  $\, \tau \, \rightarrow \, \tau/n$ ($n$ integer), as
{\em exact generators of the renormalization group}
of the two-dimensional Ising model\footnote[2]{A similar 
identification of these isogenies $\, \tau \, \rightarrow \, n \cdot \tau$
with exact generators of the renormalization group can 
be introduced for any lattice model
with an elliptic parametrization (Baxter model, ...).}.

One does not need to restrict the analysis to the real fixed points of the
transformations.
If one considers the Landen transformation 
(\ref{landenA})  as an algebraic transformation 
 of the {\em complex variable} $\, k$ and if one
solves $\, k_1^2\, -k^2=\, 0$, one obtains:
\begin{eqnarray}
\label{polk}
k \cdot (1-k) \cdot ({k}^{2} +3k +4)\,\, =\,\,\,  0.
\end{eqnarray} 
The quadratic roots 
\begin{eqnarray}
\label{polkbis}
k^2 + 3\,k +4 \, =\,\, 0,
\end{eqnarray} 
are (up to a sign)
 {\em fixed points} of (\ref{landenA}).
We thus see the occurrence of {\em additional non-trivial complex
selected values} of the modulus $\, k$, beyond the 
well-known values $k=\, 1, \, 0,\, \infty $ (corresponding to 
degeneration of the elliptic curve into a {\em rational curve}).
Physically, these well-known values $k=\, 1, \, 0,\, \infty $
correspond  to the {\em critical} Ising model ($k\, = \, 1$)
 and to (high-low temperature)
trivializations of the model ($k=\,  \, 0,\, \infty$).

\subsection{Complex multiplication for elliptic curves 
as (complex) fixed points of the
renormalization group}
\label{complex}

We come now to our point. 
The first ``unexpected''  singularities
$\, 1+3\, w \, + 4 \, w^2\, = \, \, 0$
found~\cite{ze-bo-ha-ma-04,ze-bo-ha-ma-05} for the Fuchsian linear differential 
equation of $\, \chi^{(3)}$, 
and also in other $n$-fold integrals of the Ising
class~\cite{bo-ha-ma-ze-07},
reads in the variable $k=\, s^2$ as
\begin{eqnarray}
\label{polk2}
(k^2\, +3\, k\, +4)\, (4\, k^2\, +3\, k\, +1 )\, \, = \,\,\, 0.
\end{eqnarray}
The first polynomial\footnote[2]{
Note that the two polynomials in (\ref{polk2}) are related by the
Kramers-Wannier  duality $\, k\,  \rightarrow \, 1/k$.}
 corresponds to {\em fixed points}
of the Landen transformation (see (\ref{polk})). 
In other words we see that the selected quadratic
values $\, 1+3\, w \, + 4\, w^2\, = \, \, 0$,
occurring in the (high-temperature) susceptibility of the Ising model
as singularities of the three-particle term $\, \chi^{(3)}$, 
can be seen {\em as fixed points of the renormalization group  
when extended to complex values of the modulus} $\, k$.

For elliptic curves in fields of characteristic zero,
the only well-known selected set  of values  for  $k\,$ corresponds
to the values for which the elliptic curve has {\em complex 
multiplication}~\cite{broglie}.
Complex multiplication for elliptic curves corresponds to algebraic integer 
values (integers in the case of the Heegner numbers, see \ref{d})
of the modular $\, j$-function,
which corresponds to Klein's 
absolute invariant multiplied by 
$\, (12)^3 = \,  1728$:
\begin{eqnarray}
\label{jfun}
j(k) \, = \, \, \,256 \cdot  
 {\frac { \left( 1-{k}^{2}+{k}^{4} \right)^{3}}{ k^{4} \cdot 
 \left( 1-k^{2} \right)^{2}}}.
\end{eqnarray}

A straightforward calculation of the elliptic nome (\ref{defq}) gives,
for the polynomials (\ref{polk2}), respectively,
an exact value for $\tau$, the half period 
ratio, as very simple {\em quadratic numbers}:
\begin{eqnarray}
\label{tauvalue}
\tau_1 \, \, = \, \, \, {{ \pm 3\, +i \, \sqrt{7} } \over {4}},
\qquad \quad
\tau_2 \, \, = \, \, \, {{\pm 1\, +i \, \sqrt{7} } \over {2}} 
\end{eqnarray}
These quadratic  numbers actually correspond to 
{\em complex multiplication} of the elliptic curve and for both
one has $j \, =\, (-15)^3$. These two quadratic numbers 
are such that $\, 2\,\tau_1 \mp 1 =\tau_2$. 
Let us focus on $\tau_2$ for which we can write:
\begin{eqnarray}
\label{quadratau}
\tau \,  = \,\, \, 1\, -{{2} \over {\tau}}.
\end{eqnarray}

Taking into account the two modular group involutions
 $\tau\,\,  \rightarrow  \,\, 1\, -\tau$
and $\,\tau \,\,  \rightarrow \,\, 1/\tau$, 
we find that $1\, -\, 2/\tau$ is, {\em up to the modular group},
equivalent to $\, \tau /2$.
The quadratic relation $\, \tau^2\,  - \tau\, +2\, = \, \, 0 $ thus
amounts to looking at the fixed
 points of the Landen transformation
$\, \tau \, \rightarrow \, 2\, \tau$ {\em up to the modular group}.
This is, in fact a quite general statement.
The {\em complex multiplication} values can all be seen as fixed
points,  {\em up to the modular group},
of the generalizations of Landen transformation,
namely  $\, \tau \, \rightarrow \, n\, \tau$ 
for $\, n$ integer, $\tau^2\,  - \tau\, +n\, =  \, 0$ or
$\tau \,  = \, 1\, -{{n} \over {\tau}} \,\simeq  \, n \cdot \tau$,
where $\simeq$ denotes the equivalence {\em up to the modular group}.

\ref{c} presents an alternative view by considering the solutions as
fixed points under Landen transformations of the modular $j-$function.

In view of the remarkable mathematical (and physical)
 interpretation of the quadratic
values $\, 1+3\, w \, + 4\, w^2\, = \, 0$
in terms of {\em complex multiplication for elliptic curves, or fixed points of the
renormalization group}, it is natural to see 
if such a ``complex multiplication of elliptic curves'' interpretation
also exists for other singularities of $\, \chi^{(n)}$, and 
as a first step, for the singularities of the linear differential equations 
of our $\, n$-fold  integrals (\ref{In}), that 
we expect to be identical, or at least
 have some overlap, with the singularities of the $\, \chi^{(n)}$. 

Noting that the modular $j-$function is a function of 
$\, s^2$  or $\, w^2$ (see (\ref{jinw}) in \ref{d})
the occurence of  $\, 1+3\, w \, + 4\, w^2\, = \, 0$ as a selected 
quadratic polynomial condition means, at the same time, the occurence
of the other quadratic polynomial condition $\, 1-3\, w \, + 4\, w^2\, = \, 0$
(see \ref{d} and \ref{c8}).  

Besides  $\, 1-3\, w \, + 4\, w^2\, = \, 0$,
we have found two other polynomial conditions
which correspond to remarkable integer values of the  modular $\, j$-function.
The singularities $\, 1\, -8\, w^2 =\, 0\, $ correspond to 
$j=\, (12)^3$  and $\tau=\pm 1 +i$ (see \ref{d}). They correspond to
``Nickelian singularities'' for $\, \chi^{(8)}$ (and thus $\Phi^{(8)}_H$)
and to ``non-Nickelian singularities'' for  $\Phi_H^{(10)}$.
Another polynomial condition is  $\, 1\, -32\, w^2=\, 0\, $, which gives
``non-Nickelian singularities'' that begin 
to appear at $n=10$ for $\, \Phi_H^{(10)}$.
These singularities correspond to the integer value of
the modular $\, j$-function, $j=\, (66)^3$ and 
to $\tau=\, 2\, i$ or  $\tau =\, -4/5\, +\, 2\, i/5$.

\subsection{Beyond elliptic curves}
\label{beyond}

Among the singularities of the linear ODE for $\, \Phi_H^{(n)}$ given in
 (\ref{singphi5b}), (\ref{singphi6}) or obtained from the formula
given in \ref{singLand} up to $n=\, 15$, we have found no other 
singularity identified with
selected algebraic values of the
 modular $\, j$-function  corresponding to 
complex multiplication for elliptic curves.
Could it be that the (non-Nickelian) singularities
(\ref{singphi5}), (\ref{singphi6}), which
 do not match with complex multiplication
 for elliptic curves, are actually remarkable selected situations for 
mathematical structures more complex than elliptic curves ?
With these new singularities, 
 are we possibly exploring some remarkable ``selected 
situations'' of some {\em moduli space of curves 
corresponding to pointed (marked) curves}~\cite{Barad}, instead
of simple elliptic curves~\cite{Harris}?
In practice this just corresponds to considering
a product of $\, n$ times a rational, or elliptic, curve
 minus some sets of remarkable codimension-one
algebraic varieties~\cite{Diag}, $\, x_i\, x_j\, = \, 1$, 
 $\, x_i\, x_j\, x_k\, = \, 1$, 
hyperplanes $\, x_ i \, = \, x_j$, $\cdots$ 

We try to fully understand the singularities
of the $\, n$-fold integrals 
corresponding to the $\, \chi^{(n)}$, that is to say 
particular  $\, n$-fold integrals linked 
to the theory of elliptic curves. These $\, n$-fold integrals
are more involved than the (simpler)
$\, n$-fold integrals introduced by
Beukers, Vasilyev~\cite{Va1,Va2} and Sorokin~\cite{So2,So3},
or the Goncharov-Manin integrals~\cite{Fischler} which occur 
in some {\em moduli space of curves}~\cite{Bloch,Goncharov}
simply corresponding to a 
{\em product of  rational curves}
 ($CP_1\, \times CP_1\ \cdots \,  \times   CP_1 $).  
An example of such integrals, linked\footnote[5]{Note that  $\, \zeta$
(or the polyzeta) function evaluated at integer 
values ($\zeta(3)$, $\, \zeta(5)$, ...) 
do occur in our more involved $\, n$-fold integrals, in particular in the
representation of the connection matrices~\cite{ze-bo-ha-ma-05c}
 of the differential Galois group 
of the  Fuchsian linear ODEs of  $\, \chi^{(n)}$.} to $\, \zeta(3)$,
 is displayed\footnote[8]{
These  $\, n$-fold 
integrals~\cite{Fischler3, Brown, Huttner, Racinet, Fischler2}
 look almost the same as the ones 
we have introduced and analyzed in the study of the 
diagonal susceptibility of the Ising model~\cite{Diag}
for which $\, n$-th root of unity singularities occur.} in \ref{G}.

\vskip 0.1cm

Let us close this section by noting that Heegner numbers and,
 more generally, {\em complex
multiplication} have already occurred in 
other contexts, even if the statement
was not explicit. In the framework of
 the construction of Liouville field theory,
Gervais and Neveu have suggested~\cite{Gervais} 
new classes of critical statistical models, where, besides
 the well-known $\, N$-th root of unity situation, 
they found the following selected values
of the {\em multiplicative crossing} $t$ ~\cite{Rammal2}:
\begin{eqnarray}
\label{othervaluesint}
&&t \, = \, \,\, e^{ i\, \pi \, (1+i \sqrt{3})/2}
 \, = \, \,\, i \cdot e^{-\pi\, \sqrt{3}/2 }, \\
\label{othervaluesint2}
&&t \, = \, \,\, e^{ i\, \pi \, (1+i)}
 \, = \, \, -e^{-\pi}.
\end{eqnarray}
 If one wants to see  this multiplicative crossing as 
 a modular nome,  
the two previous situations actually correspond to
selected values of the modular $\, j$-function namely  
$\,j((1+i \sqrt{3})/2) \, = \, \, (0)^3$ for (\ref{othervaluesint}), 
and $\,j(1+i) \, = \, \, (12)^3$ for (\ref{othervaluesint2}), which 
actually correspond to {\em Heegner numbers
and, more generally, complex multiplication}~\cite{broglie}.
 It is however important not to
feed the confusion already too prevalant
 in the literature, between 
 a {\em ``temperature-like'' nome} such as (\ref{defq}) 
and a {\em multiplicative crossing modular nome}.
In the Baxter model~\cite{Baxter,Baxter2}, the first is denoted by $\, q$ and the
second one by $\, x$.
In fact one probably has, {\em not one, but two modular groups} taking place,
one acting on the ``temperature-like'' nome $\, q$ 
 and the other acting on the multiplicative 
crossing $\, x$. We will not go further along this quite speculative line 
which amounts to introducing {\em elliptic quantum groups}~\cite{Mano}
and  {\em elliptic gamma functions}\footnote[1]{Which can be seen~\cite{Varchenko}
 as ``automorphic forms of degree 1''
 when the Jacobi modular forms are ``automorphic forms of degree 0''
and are associated (up to simple semi-direct products)
with $\, SL(3, \, Z)$ instead of  $\, SL(2, \, Z)$} (generalization
 of theta functions\footnote[9]{The partition function of the Baxter model 
can be seen as a ratio and product of 
elliptic gamma functions and theta functions.
It  is thus naturally expressed as a double infinite 
product. Similar double, and even triple, products
 appear in correlation functions of
 the eight vertex model~\cite{JMN,JKKMW}.}).
 
\section{Conclusion}
\label{conclu}

The ultimate goal of our ``Ising class'' 
integrals is to 
get some insight into the $\, \chi^{(n)}$ and, hopefully, into the 
susceptibility of the Ising model. For that purpose
we have introduced $\, n$-fold integrals (\ref{In}) such that 
we expect the singularities of the corresponding linear ODE
to overlap, as much as possible, with the singularities 
of the linear ODE for the  $\, \chi^{(n)}$. We have obtained
 the linear differential equations
 for these $\, n$-fold integrals $\, \Phi^{(n)}_H$,  
 up to $\, n=4$ and up to $\, n=6$ modulo a prime.
From these exact results together with an exhaustive
 Landau singularity analysis, we provided a quite complete description of 
the singularities of these linear ODEs.

From the Landau conditions, the singularity structures are explained.
The singularities corresponding to $\, \Phi_H^{(n)}$ are found to also occur
at a higher predefined order $\, p\, > \, n$.
With these multiple integrals and the associated Landau
conditions, we have been able to understand why the simple
integrals $\, \Phi_D^{(n)}$ have succeeded reproducing the Nickelian
singularities and the new quadratic $\, 1 +3w +4w^2=\, 0$.
These simple integrals
appear to be "a first approximation" to $\Phi_H^{(n)}$. Other
 one-dimensional integrals pop up to account for the additional singularities
not occurring for $\, \Phi_D^{(n)}$.

We have then a remarkable finding that, the singularities for the multiple
integrals can be associated
 with the singularities for a finite number of one
dimensional integrals.
If the singularities, associated with these $\, n$-fold integrals (\ref{In}),
happen to be identical with (or to overlap)
the singularities associated with the $\, \chi^{(n)}$,
it becomes important to understand this  mechanism for the
$\, \chi^{(n)}$ themselves.
If this mechanism of singularity embedding occurs
for $\, \chi^{(n)}$, it might be explained by a Russian doll structure
for the same linear differential operators. We know that the linear 
differential operator for $\, \chi^{(1)}$ (respectively $\, \chi^{(2)}$)
is ``contained''  in (rightdivides) the linear differential operator for $\, \chi^{(3)}$
(respectively $\, \chi^{(4)}$), and furthermore we even have
direct sum decomposition properties. For the $\Phi_H^{(n)}$, it is not
these mechanisms which are at work.

Our primary goal in this study is to identify as many singularities as possible
 for the $\, \chi^{(n)}$.
The singularities of the ODEs associated with the
 $\, \Phi_H^{(n)}$ quantities correspond, in the Landau
equations  framework, to {\em leading pinch singularities} (relatively
to the singularities manifolds $\, D(\phi,\, \psi)=\, 0$). For the 
other quantities previously studied~\cite{bo-ha-ma-ze-07}  which belong  
to the Ising class integrals, the same feature holds.

At this step, the natural questions arising are: 
whether the  scheme, from the Landau singularities point of
view, which holds for $\, \Phi_H^{(n)}$,  still holds
for $\, \chi^{(n)}$ and whether the singularities of 
$\, \Phi_H^{(n)}$ can be considered as singularities of the
$\, \chi^{(n)}$ ?

From the Landau singularities viewpoint, the Fermionic
determinant $\, G(n)^2$ is going to introduce new manifolds
of singularities. When the Lagrange multipliers
relative to the singularities manifolds introduced by
the Fermionic determinant are all set equal to zero, one
deals with the Landau equations of the $\, \Phi_H^{(n)}$
quantities.
Thus the singularities obtained for the  $\, \Phi_H^{(n)}$
quantities are also solutions of the Landau equations
of the $\, \chi^{(n)}$.
However this feature does not mean that the
singularities of the  $\, \Phi_H^{(n)}$ quantities will
necessarly appear as singularities
of the $\, \chi^{(n)}$ ODEs. Indeed some selection rules may take
 place and may reject some of them. For instance,
 one expects singularities linked to the $\, \prod y_i$
to occur for the Landau singularities of the
 $\, \Phi_H^{(n)}$. One finds that some selection rules exclude them.  
Our ``educated guess'' is that all the Landau singularities of the $\, \Phi_H^{(n)}$
will be in the  Landau singularities 
of the  $\, \chi^{(n)}$, however we do not exclude
the possibility that the $\, \chi^{(n)}$ will have more  Landau singularities 
than the $\Phi_H^{(n)}$. Another ``educated guess'' is that the  
Landau singularities of the  $\, \chi^{(n)}$ will exhibit a similar 
embedding that the one we found for the $\Phi_H^{(n)}$.
This naturally raises the question already considered in~\cite{ze-bo-ha-ma-05b},
of a ``strong'' Russian doll structure for the linear differential operators 
of the  $\, \chi^{(n)}$, namely that the linear differential operator
of $\, \chi^{(3)}$ (resp. $\, \chi^{(4)}$) could 
right-divide the linear differential operator
of $\, \chi^{(5)}$ (resp. $\, \chi^{(6)}$), and so on.

This knowledge of the singularities will help
 in the search for the corresponding linear ODE.
For instance, we have 24  head polynomial ``candidates'' for  $\, \chi^{(5)}$ and
19 ``candidates'' for  $\, \chi^{(6)}$ that can, from the outset, 
 be put  in front of the higher order derivative of the unknown linear ODE.
From the knowledge we have gained from all these $\, n$-fold integrals
of the ``Ising class'', one
can guess the order of magnitude of the multiplicity of some
singularities. Furthermore, as shown for the linear ODE for $\, \Phi_H^{(5)}$
and $\, \Phi_H^{(6)}$ (and also from previous ODEs), we know that the
``cost'' (in terms of the number of series coefficients) will be much less for
a non minimal order linear ODE than for the minimal order one.

Concerning the non Nickelian singularities that the multiple integrals
$\,\Phi_H^{(n)}$ have given, we focussed on
 $\, 1+3\, w \, + 4 \, w^2\, = \, \, 0$
which actually occurs for the linear ODE of $\, \chi^{(3)}$,
or for  $\, \chi^{(3)}$ seen as a function of $\, s$.
As far as a {\em mathematical interpretation} is concerned,
we have shown that this quadratic polynomial condition corresponds to a
selected situation for elliptic curves
namely the {\em occurrence of complex multiplication}.
The other  non-Nickelian (candidate) singularities,
(\ref{singphi5}), (\ref{singphi6}) {\em do not correspond to}
 complex multiplication of elliptic curves.

Assuming that the non Nickelian singularities obtained in the linear ODE
for the integrals (\ref{In}), will be, at least, included in 
 those for the $\, \chi^{(n)}$, 
various lines of thought are possible.

One may imagine that the decomposition of the susceptibility of the
Ising model in terms of an infinite sum of  $\, \chi^{(n)}$ 
is quite an artificial one with no deep mathematical meaning, i.e.
$\,\chi^{(n)}$ are quite arbitrary $\, n$-fold integrals.
In this case, no interpretation within the theory of elliptic curves
has to be looked for and  the occurrence for $\, 1+3\, w \, + 4 \, w^2\, = \, 0$
of complex multiplication
for elliptic curves would be just a  coincidence.

Another option amounts to saying that one needs to introduce (motivic)
mathematical structures~\cite{Fischler3, Brown, Huttner, Racinet, Fischler2}
 {\em beyond the theory of elliptic
 curves} (moduli spaces, marked curves, ...), and
beyond the elliptic curves  of the Ising (or Baxter) model,
to get a {\em mathematical interpretation of these singularities}.
We tend to favour the latter option.

\vskip .3cm

\vskip .3cm

\vskip .3cm

\textbf{Acknowledgments:}
 We have derived great benefit from discussions on various 
aspects of this work with F. Chyzak, S. Fischler, P. Flajolet,
A. J. Guttmann, L. Merel, B. Nickel,  I. Jensen,  B. Salvy and J-A. Weil.
 We thank A. Bostan for a search of linear 
ODE mod. prime with one of his Magma programs. 
We acknowledge  CNRS/PICS financial support. 
 One of us (NZ) would like to acknowledge the kind hospitality
at the LPTMC where part of this work has been completed.  
One of us (JMM) thanks  MASCOS (Melbourne) where
 part of this work was performed.

\vskip .3cm
\pagebreak

\vskip .3cm 

\appendix

\section{Series expansions of $\Phi_H^{(n)}$ and of single
integrals  $\Phi_k^{(n)}$}
\label{a0}

We give in this Appendix, the series expansion that has been used
for $\Phi_H^{(n)}$. Expanding the integrand of (\ref{In}) in the variables
$x_j$, one obtains
\begin{eqnarray}
\label{integral}
\Phi_H^{(n)} \, \,= \,\,\,\, \,  {\frac{1}{n!}}  \cdot 
 \prod_{j=1}^{n-1} \int_0^{2\pi} {\frac{d\phi_j}{2\pi}}  
  \cdot  
\sum_{p=0}^\infty \, (2-\delta_{p,\, 0}) \cdot   \prod_{j=1}^{n}\,y_j \, x_j^p. 
\end{eqnarray}

We make use of the $y_j\, x_j^p$ Fourier expansion
\cite{ze-bo-ha-ma-04, ze-bo-ha-ma-05, ze-bo-ha-ma-05b}
\begin{eqnarray}
\label{fourier}
y_j\, x_j^p \,\, =\,\,\,  
w^p \cdot  \sum_{k=-\infty}^\infty \, w^{\vert k \vert}\cdot 
a(p, \vert k \vert)\cdot 
Z_j^k, \qquad Z_j=\exp(i\, \phi_j)
\end{eqnarray}
where $\, a(k,p)$ is a {\em non-terminating}
hypergeometric function that  reads (with $m=k+p$):
\begin{eqnarray}
&&a(k,p) \, =\,\, 
\, {{m }\choose {k}} \times  \\
&& \quad \quad {_4}F_3 \Bigl( {\frac{1+m}{2}},
 {\frac{1+m}{2}},
{\frac{2+m}{2}}, {\frac{2+m}{2}}; 1+k, 1+p, 1+m; 16\, w^2 \Bigr).
\nonumber 
\end{eqnarray}
We define  $\, \langle  \rho \rangle$ by
\begin{eqnarray}
 \langle  \rho \rangle \,\,  =\, \, \, \,  
 \, \Bigl( \prod_{j=1}^{n} \int_0^{2\pi} {\frac{d\phi_j}{2\pi}} \Bigr)
  \cdot  2\pi\, \delta \left(\sum_{j=1}^n\, \phi_j \right)\cdot \rho 
\end{eqnarray}
where the angular constraint is introduced through the delta function
that has the Fourier expansion:
\begin{eqnarray}
2\pi\, \delta \left(\sum_{j=1}^n\, \phi_j \right)\, 
\,=\,\,\, \,  \sum_{k=-\infty}^\infty
\left( Z_1\, Z_2 \, \cdots Z_n \right)^k  
\end{eqnarray}
The integrals (\ref{integral}) become
\begin{eqnarray}
\Phi_H^{(n)} \, \,= \,\,\,\,\,   {\frac{1}{n!}}  \cdot 
\sum_{k=-\infty}^\infty \,   
\sum_{p=0}^\infty \,(2-\delta_{p,\, 0})\cdot 
  \langle \prod_{j=1}^{n}\,y_j \, x_j^p\, Z_j^k \rangle
\end{eqnarray}
where the integration is over independent angles.

Using the Fourier expansion (\ref{fourier}), one obtains
the integration rule
\begin{eqnarray}
 \langle  y_j\,x_j^p\, Z_j^k \rangle \,\,  =\,\,\,\, 
 w^{p+\vert k \vert}\cdot  a\left(p, \vert k \vert \right)
\end{eqnarray}
and finally:
\begin{eqnarray}
\Phi_H^{(n)}\, =\, \, \,\,  \, {\frac{1}{n!}}  \cdot
\sum_{k=0}^{\infty} \sum_{p=0}^{\infty}\,
(2-\delta_{k,\, 0}) \cdot (2-\delta_{p,\, 0}) \cdot  w^{n(k+p)} \cdot a^n(k,p).
\end{eqnarray}

\vskip 0.1cm

The derivation of the series expansions for the one dimensional integrals
(\ref{diag-gener}) proceeds along similar lines.
The integrand of the integrals (\ref{diag-gener}) is expanded in $x$
\begin{eqnarray}
\label{onedim}
{{1} \over {  1\,\, -x^{n-k}(\phi)  \cdot  x^k (  \phi_n )  }}
\, =\,\, \, \,\sum_{p=0}^\infty \, x^{p(n-k)} (\phi)\, x^{p\,k} (\phi_n)
\end{eqnarray}
with
\begin{eqnarray}
\phi_n \, =\,\,  \,- {\frac{n-k}{k}} \cdot \phi \,\,  + {2 \pi j \over k}.
\end{eqnarray}
Here, we use the Fourier expansion
\begin{eqnarray}
x^m \,\, = \,\,\,\,  w^m \cdot   
\sum_{p=0}^{\infty} (2-\delta_{p,\, 0}) 
 \cdot w^{p}\cdot  b(p, m) \cdot \cos(p\,\phi) 
\end{eqnarray}
where $\, b(k,p)$ is a {\em non-terminating} hypergeometric function
that  reads (with $m=k+p$):
\begin{eqnarray}
&&b(k,p) \,\,  =\,\, \,\, 
\, {{m-1 }\choose {k}} \times  \\
&& \quad  \quad \quad {_4}F_3 \Bigl( {\frac{1+m}{2}},
 {\frac{1+m}{2}},
{\frac{2+m}{2}}, {\frac{m}{2}}; 1+k, 1+p, 1+m; 16\, w^2 \Bigr).
\nonumber 
\end{eqnarray}

The integration of the one-dimensional integrals (\ref{onedim}) gives
\begin{eqnarray}
&& \Phi_k^{(n)}\, \,  =\, \,\, \, \, \, 
 \langle {{1} \over {  1\,\, -x^{n-k}(\phi) 
 \cdot  x^k (\phi_n) }} \rangle \,=\, \, 
 \sum_{p=0}^\infty \,
 \sum_{p_1=0}^\infty \, \sum_{p_2=0}^\infty \,
 (2-\delta_{p_1,\, 0})  \cdot (2-\delta_{p_2,\, 0})   \nonumber \\
&& \qquad \quad \times \, w^{p n +p_1+p_2} \cdot 
 b\left( p_1, p\,(n-k) \right)
 \cdot   b\left( p_2, p\,k \right)\, I(p_1,p_2)
\end{eqnarray}
with
\begin{eqnarray}
I(p_1, p_2) \,=\,\,\, 
{1 \over 2} \cdot (1+\delta_{p_1,\, 0}) 
 \cdot  \cos \left( c \right),
\quad \qquad \quad \quad \quad \, \, \,   {\rm for}\,\,\,\,  \, 
p_2\cdot  (n-k)\,=\,\, k \cdot p_1,  \nonumber 
\end{eqnarray}
and 
\begin{eqnarray}
I(p_1, p_2) \, =\,\,\, {1 \over \pi} \, {b^2 \over b^2-p_1^2} 
\cdot  \sin (b\,\pi) \cdot  \cos (b\,\pi-c ),
\quad \quad \, \,  
{\rm for}\,\,\,\,  \, \,p_2 \cdot  (n-k)\,\,\ne\,\,k \cdot p_1,  \nonumber
\end{eqnarray}
where
\begin{eqnarray}
 b\,=\,\, {n-k \over k}\cdot  p_2, \qquad \quad \quad
 c\,=\, \, {2 \pi\, j \over k}\cdot p_2.
\end{eqnarray}

\section{Linear differential equations of some $\Phi_H^{(n)}$}
\label{a}

\subsection{Linear  ODE for  $\, \Phi_H^{(3)}$}

The minimal order linear differential equation satisfied by 
$\Phi_H^{(3)}$ reads
\begin{eqnarray}
\sum_{n=0}^{5}\,a_{n}(w)\cdot
 {\frac{{d^{n}}}{{dw^{n}}}}F(w)\,\,=\,\,\,\,\,0,
\end{eqnarray}
where
\begin{eqnarray}
&&a_5 (w) \,  = \, 
\left(1-w \right)  \left(1-4\,w \right)^{4}
 \left( 1+4\,w \right)^{2} 
\left( 1+2\,w \right)  \nonumber \\
&&\qquad \times  (1+3\,w+4\,{w}^{2}) \cdot  {w}^{3} 
\cdot  P_5(w), \\
&&a_4 (w) \,  = \, 
\left(1-4\,w \right)^{3} \left( 1+4\,w \right) \cdot  {w}^{2}  
 \cdot  P_4(w), \qquad \nonumber  \\
&&a_3 (w) \, =\,
-2\, \left(1-4\,w \right)^{2} \cdot w 
 \, P_3(w), \qquad
a_2 (w) \, = \, 2\, (1-4\,w)  \cdot  P_2(w), 
\qquad \nonumber \\
&&a_1 (w)  =  \,-8\, P_1(w), \qquad \quad
a_0 (w)  = \, -96\, P_0(w), \nonumber 
\end{eqnarray}
with
\begin{eqnarray}
&&P_5(w) \, =\, 
-5+21\,w+428\,{w}^{2}+5364\,{w}^{3}-82416\,{w}^{4}
-299504\,{w}^{5}
 \nonumber \\
&&\quad\quad +714944\,{w}^{6} +3127872\,{w}^{7}-8220672\,{w}^{8}
-25858048\,{w}^{9}
\nonumber \\
&& \quad\quad -7077888\,{w}^{10}  
 +31424512\,{w}^{11}-42467328\,{w}^{12}
\nonumber \\
&& \quad\quad -31457280\,{w}^{13}-4194304\,{w}^{14}+4194304\,{w}^{15}, 
 \nonumber \\
&&P_4(w)\, =\,
-40+7\,w+5232\,{w}^{2}+37159\,{w}^{3}-447778\,{w}^{4}
-4947500\,{w}^{5}\nonumber \\
&&\quad \quad+19493448\,{w}^{6} 
+258464112\,{w}^{7}+499205984\,{w}^{8}-1612751808\,{w}^{9}\nonumber \\
&&\quad\quad
-4667817856\,{w}^{10}  +13827459072\,{w}^{11}+67078416384\,{w}^{12}
\nonumber \\
&&\quad \quad
+62392041472\,{w}^{13}-81535369216\,{w}^{14} 
-116835483648\,{w}^{15}\nonumber \\
&&\quad \quad+124662054912\,{w}^{16}
+146016305152\,{w}^{17}-197258117120\,{w}^{18} \nonumber \\
&& \quad\quad -131667591168\,{w}^{19}-11676942336\,{w}^{20}
+15032385536\,{w}^{21},
  \nonumber \\
&&P_3(w) \, =\,\, 
35\, -25\,w-8683\,{w}^{2}\, -10149\,{w}^{3}+619246\,{w}^{4}
+5273820\,{w}^{5}\nonumber \\
&& \quad\quad
-52472072\,{w}^{6}  -588147792\,{w}^{7} 
 +491073248\,{w}^{8}+18721819584\,{w}^{9}\nonumber \\
&& \quad \quad +47622771584\,{w}^{10} 
 -97459630592\,{w}^{11}-441418588160\,{w}^{12}\nonumber \\
&& \quad \quad
+651003559936\,{w}^{13}
 +4694018588672\,{w}^{14}+4729946636288\,{w}^{15}\nonumber \\
&& \quad \quad
-7193770328064\,{w}^{16} 
 -11814519701504\,{w}^{17}+7399599505408\,{w}^{18}\nonumber \\
&& \quad \quad
+10981996494848\,{w}^{19} 
 -16439524196352\,{w}^{20}-10434623045632\,{w}^{21}\nonumber \\
&& \quad \quad
-916975517696\,{w}^{22} +1125281431552\,{w}^{23},
 \nonumber \\
&&P_2 (w) \, =\, 
-10+101\,w+11088\,{w}^{2}-42855\,{w}^{3}
-1117278\,{w}^{4}-1918516\,{w}^{5}\nonumber \\
&& \quad\quad
+72221464\,{w}^{6}  +460080656\,{w}^{7} -4999186016\,{w}^{8}
\nonumber \\
&&\quad \quad
-33474428224\,{w}^{9}+67440200320\,{w}^{10} 
 +808560558592\,{w}^{11}\nonumber \\
&&\quad \quad +535166693376\,{w}^{12}
-6771457933312\,{w}^{13}-7468556451840\,{w}^{14} \nonumber \\
&&\quad \quad +46143514476544\,{w}^{15}
+91488863125504\,{w}^{16} \nonumber \\
&&\quad \quad-75107733078016\,{w}^{17}
-239438663778304\,{w}^{18}+31904728350720
\,{w}^{19} \nonumber \\
&& \quad \quad +234058806198272\,{w}^{20}-237446193217536\,{w}^{21}
-164181567340544\,{w}^{22} \nonumber \\
&& \quad \quad -18975165513728\,{w}^{23}+16973710753792\,{w}^{24}, 
 \nonumber \\
&&P_1 (w) \, =\, 
-5-1142\,w+8106\,{w}^{2}+210846\,{w}^{3}
-1070376\,{w}^{4}-7771160\,{w}^{5}\nonumber \\
&&\quad
-22029952\,{w}^{6}  +833894752\,{w}^{7}
+3334510976\,{w}^{8}-39736449920\,{w}^{9}\nonumber \\
&&\quad
-156101859328\,{w}^{10} +663306718208\,{w}^{11}
+2995615555584\,{w}^{12}\nonumber \\
&& \quad
-5033154314240\,{w}^{13}  -26250785980416\,{w}^{14}
+28618066755584\,{w}^{15}\nonumber \\
&& \quad
+158047775227904\,{w}^{16} 
 -42836217036800\,{w}^{17}-410317620248576\,{w}^{18}\nonumber \\
&& \quad -95925074657280\,{w}^{19}
 +462245318361088\,{w}^{20}-328990199906304\,{w}^{21}\nonumber \\
&& \quad
-249443110617088\,{w}^{22} 
 -35270271434752\,{w}^{23}+24464133718016\,{w}^{24},
 \nonumber \\
&&P_0 (w) \, =\, 
 -5+58\,w+3234\,{w}^{2}-18994\,{w}^{3}-229330\,{w}^{4}
 +1516\,{w}^{5} \nonumber \\
&&\quad
+7017504\,{w}^{6}  +74689472\,{w}^{7}
-647069792\,{w}^{8}-4260373952\,{w}^{9} \nonumber \\
&&\quad
+15887163648\,{w}^{10} +96789618688\,{w}^{11}
-136120508416\,{w}^{12} \nonumber \\
&&\quad -917765144576\,{w}^{13} +877996605440\,{w}^{14}
+5646695006208\,{w}^{15} \nonumber \\
&&\quad-2888887697408\,{w}^{16} 
 -16785155817472\,{w}^{17}-5241017729024\,{w}^{18}\nonumber \\
&& \quad
+17952426426368\,{w}^{19}
 -13058311192576\,{w}^{20}-9329742708736\,{w}^{21} \nonumber \\
&&\quad
-1275605286912\,{w}^{22}   +824633720832\,{w}^{23}. 
\end{eqnarray}

\subsection{Linear  ODE for  $\, \Phi_H^{(4)}$}
The minimal order linear differential equation satisfied by 
$\Phi_H^{(4)}$ reads (with $x=16w^2$)
\begin{eqnarray}
\sum_{n=0}^{6}\,a_{n}(x)\cdot {\frac{{d^{n}}}{{dx^{n}}}}F(x)
\,\,=\,\,\,\,\,0,
\end{eqnarray}
where
\begin{eqnarray}
&&a_6 (x)\,   = \,   64\, \left(x-4 \right) 
 \, (1-x)^{4}{x}^{4}\cdot  P_6(x),
\quad a_5 (x) \,  = \,
 -128\, (1-x)^{3}\, {x}^{3} \cdot  P_5(x), \qquad
\nonumber \\
&&a_4 (x)\,  = \,  16\, (1-x)^{2}{x}^{2}
 \cdot P_4(x), \qquad 
a_3 (x) \, = \, -64\, (1-x) \, x \cdot  P_3(x), \qquad
\nonumber  \\
&&a_2 (x) \, = \,-4 \cdot  P_2(x), \quad
 a_1 (x) \, = \,-8\cdot  P_1(x), \quad 
a_0 (x) \,  = \,  -3\, (1-x) \cdot  P_0(x), \nonumber
\end{eqnarray}
with:
\begin{eqnarray}
&&P_6(x) \, =\,\,  
 128+2233\,x-2847\,{x}^{2} +3143\,{x}^{3}
-3601\,{x}^{4}+144\,{x}^{5}-64\,{x}^{6}, 
 \nonumber \\
&&P_5(x) \, =\, 
 3712+51523\,x-216377\,{x}^{2}+289918\,{x}^{3}
-312896\,{x}^{4}\nonumber \\
&& \quad \quad +262111\,{x}^{5} 
  -63167\,{x}^{6}+5512\,{x}^{7}-896\,{x}^{8}, 
 \nonumber \\
&&P_4(x) \, =\, 
 -121856-1102304\,x
+11038289\,{x}^{2}-26106487\,{x}^{3}\nonumber \\
&& \quad \quad +31515802\,{x}^{4}
  -31027694\,{x}^{5}+21291429\,{x}^{6}
-5166011\,{x}^{7}\nonumber \\
&& \quad \quad 
+410160\,{x}^{8}-67776\,{x}^{9},
  \nonumber \\
&&P_3(x) \, =\, 
 38144+10604\,x -4644281\,{x}^{2}
+20909702\,{x}^{3}-37890772\,{x}^{4}\nonumber \\
&& \quad \quad
+42011874\,{x}^{5} -37552559\,{x}^{6}+22474036\,{x}^{7}
-5465572\,{x}^{8}\nonumber \\
&& \quad \quad +392536\,{x}^{9}-65984\,{x}^{10},
 \nonumber \\
&&P_2 (x) \, =\, 
 163840-4162688\,x-18120152\,{x}^{2}+277110610\,{x}^{3} \nonumber \\
&& \quad \quad 
-880048289\,{x}^{4}  +1357147519\,{x}^{5}
-1395938590\,{x}^{6}+1141353668\,{x}^{7} \nonumber \\
&& \quad \quad 
-621323833\,{x}^{8}  +150842795\,{x}^{9}
-9676720\,{x}^{10}+1656512\,{x}^{11}, 
\nonumber  \\
&&P_1 (x) \, =\, 
 -366592+3113752\,x+17465700\,{x}^{2}-120658444\,{x}^{3}\nonumber \\
&&\quad \quad 
+240321805\,{x}^{4}  -259277988\,{x}^{5}+219951814\,{x}^{6}
-142314304\,{x}^{7}\nonumber \\
&&\quad \quad +42534921\,{x}^{8}
-2056040\,{x}^{9}  +435200\,{x}^{10},
 \nonumber \\
&&P_0 (x) \, =\, 
 561152-1496400\,x-13171575\,{x}^{2}+
30840556\,{x}^{3}-24381198\,{x}^{4}\nonumber \\
&&\quad \quad 
+20352948\,{x}^{5} -13268091\,{x}^{6}+309360\,{x}^{7}
-120000\,{x}^{8}.  \nonumber
\end{eqnarray}

\subsection{Linear  ODE modulo a prime for
 $\, \Phi_H^{(5)}$}
\label{subsec}

The linear differential equation of minimal order seventeen 
satisfied by 
$\Phi_H^{(5)}$ is of the form 
\begin{eqnarray} 
\label{a5}
\sum_{n=0}^{17}\,a_{n}(w)\cdot 
{\frac{{d^{n}}}{{dw^{n}}}}F(w)\,\,=\,\,\,\,\,0, 
\end{eqnarray} 
with 
\begin{eqnarray} 
&&a_{17} (w) \, = \, 
 \left( 1-4\,w \right)^{12} \left( 1+4\,w \right)^{9} 
( 1-w)^{2} 
\left( w+1 \right) \left( 1+2\,w \right) 
\nonumber \\ 
&& \times 
\left( 1+3\,w+4\,{w}^{2} \right)^{2} \left( 1-3\,w+{w}^{2} \right) 
\left( 1+2\,w-4\,{w}^{2} \right) \, \nonumber \\ 
&& \times (1+4\,w+8\,{w}^{2}) 
\left( 1-7\,w+5\,{w}^{2}-4\,{w}^{3} \right) \nonumber \\ 
&& \times \left( 1-w-3\,{w}^{2}+4\,{w}^{3} \right) 
\left( 1+8\,w+20\,{w}^{2}+15\,{w}^{3}+4\,{w}^{4} \right) \cdot {w}^{12}
\cdot P_{17}(w), \nonumber \\ 
&&a_{16} (w) \, = \, 
{w}^{11} \left( 1-4\,w \right)^{11} \left( 1+4\,w \right)^{8} 
( 1-w) 
\left( 1+3\,w+4\,{w}^{2} \right) 
\cdot P_{16}(w), \nonumber \\ 
&&a_{15} (w) \, = \, 
{w}^{10} \left( 1-4\,w \right)^{10} \left( 1+4\,w \right)^{7} 
\cdot P_{15}(w), \nonumber \\ 
&&a_{14} (w) \, = \, 
{w}^{9} \left( 1-4\,w \right)^{9} \left( 1+4\,w \right)^{6} 
\cdot P_{14}(w), \nonumber \\ 
&&\cdots \nonumber 
\end{eqnarray} 
where the $ 430$ roots of $\, P_{17}(w)$ are {\em apparent singularities}. 
The degrees of these polynomials $\, P_{n}(w)$
 are such that the degrees of $a_i(w)$ are decreasing as:
$\, deg(a_{i+1}(w))=\, deg(a_i(w))\, +1$.
In fact, with 2208 terms we have found the ODE of
$\, \Phi_H^{((5)}$ at order $\, q=\, 28$ using the following ansatz for
the  linear ODE search ($Dw$ denotes $d/dw$)
\begin{eqnarray} 
\sum_{i=0}^{q} \, s(i)\cdot p(i) \cdot Dw^i
\end{eqnarray} 
with:
\begin{eqnarray} 
s(i)\,=\,\, w^{\alpha(-1 + i)} 
\cdot (1 - 16\, w^2)^{\alpha(-1 + i)} \cdot s_0^{\alpha(1 + i- q)}
\end{eqnarray} 
where $\, \alpha(n)\,=\,\, Min(0, n)$ and 
\begin{eqnarray}
&&s_0\, =\,\, (1 + w)\cdot (1 - w) \cdot (1 + 2\, w)
\cdot (1 - 3\, w + w^2)\, 
(1 +2\, w - 4\, w^2) \nonumber \\
&&\times (1 + 3\, w + 4\, w^2)\cdot (1 + 4\, w + 8\, w^2)\cdot 
(1 - 7\, w + 5\, w^2- 4\, w^3) \nonumber \\
&& \times (1 - w - 3\, w^2 + 4\, w^3)
\cdot (1 + 8\, w + 20\, w^2 + 15\, w^3 +4\, w^4)
\nonumber
\end{eqnarray} 
the $\, p(i)$ being the unknown polynomials.

The minimal order ODE is deduced from the set of
linear independant ODEs found at order 28.

\subsection{Linear  ODE modulo a prime for
  $\, \Phi_H^{(6)}$}
\label{subsec6}

The linear differential equation of minimal order twenty-seven 
satisfied by 
$\Phi_H^{(6)}$ reads (with $x=w^2$) 
\begin{eqnarray} 
\label{a6}
\sum_{n=0}^{27}\,a_{n}(x)\cdot {\frac{{d^{n}}}{{dx^{n}}}}F(x) 
\,\,=\,\,\,\,\,0, 
\end{eqnarray} 
with 
\begin{eqnarray} 
&&a_{27} (x)\, = \, 
\left( 1-16\,x \right)^{16} 
\left( 1-4\,x \right)^{3} 
\left( 1-x \right) 
\left( 1-25\,x \right) 
\left( 1-9\,x \right) {x}^{21} 
\nonumber \\ 
&& \quad \quad \times \left( 1-x+16\,{x}^{2} \right) 
\left( 1-10\,x+29\,{x}^{2} \right) 
\cdot P_{27}(x), \qquad 
\nonumber \\ 
&&a_{26} (x)\, = \, 
\left( 1-16\,x \right)^{15} 
\left( 1-4\,x \right)^{2}\,  {x}^{20} 
\cdot P_{26}(x), \qquad 
\nonumber \\ 
&&a_{25} (x)\, = \, 
\left( 1-16\,x \right)^{14} 
\left( 1-4\,x \right)\,  {x}^{19} 
\cdot P_{25}(x), \qquad 
\nonumber \\ 
&&a_{24} (x)\, = \, 
\left( 1-16\,x \right)^{13}\,  {x}^{18} 
\cdot P_{24}(x), \qquad 
\nonumber \\ 
&& \cdots 
\end{eqnarray} 
where the $307$ roots of $P_{27}(x)$ are {\em apparent singularities}. 
The degrees of the $P_n(w)$ polynomials are such that
the degrees of $a_i(w)$ are decreasing as:
$deg(a_{i+1}(w))\, =\, deg(a_i(w))+1$

In fact, with 1838 terms we have found the linear ODE of
$\, \Phi_H^{(6)}$ at order $q=\, 42$
 using the following ansatz for
the  linear ODE search ($Dx$ denotes $d/dx$)
\begin{eqnarray} 
\sum_{i=0}^{q}\, s(i) \cdot p(i) \cdot Dx^i
\end{eqnarray} 
with:
\begin{eqnarray} 
s(i)\, =\, \, x^{\alpha(-1 + i)}\cdot 
(1 - 16\, x)^{\alpha(-1 + i)} \cdot s_0^{\alpha(1 + i -q)}
\end{eqnarray} 
where $\, \alpha(n)\, =\,\,  Min(0, n)$ and
\begin{eqnarray} 
&&s_0\, =\, \, (1 - 25\, x)\cdot (1 - 9\, x)
\cdot (1 - 4\, x)\cdot (1 - x)\nonumber \\
&&\qquad \times (1 - x +16\, x^2)\cdot (1 - 10\, x + 29\, x^2)
\end{eqnarray} 
the $\, p(i)$ being the unknown polynomials.

The minimal order ODE is deduced from the set of
linear independant ODEs found at order 42.

\section{Singularities in the linear ODE for $\Phi_H^{(7)}$ and
$\Phi_H^{(8)}$}
\label{singphi7phi8}
For $\Phi_H^{(7)}$, we generated large series, (1250 coefficients
and 20000 coefficients modulo
primes), 
unfortunately, insufficient
to obtain the corresponding linear ODE.
However, by steadily increasing the order $\, q$ of the ODE 
(and consequently decreasing the degrees $\, n$
of the polynomials in front of the derivatives), one may recognize, in
floating point form, the singularities of the ODE as the roots of the
polynomial in front of the higher derivative.
A root is considered as singularity of the still unknown linear ODE, when
as $\, q$ increase (and consequently decreasing $\, n$), it
 persists with more stabilized digits.

Using $1250$ terms in the series for $\, \Phi_H^{(7)}$, the following
singularities are recognized
\begin{eqnarray}
&& \left( 1-4\,w\right)  \left( 1-5\,w+6\,{w}^{2}-{w}^{3} \right) 
 \left( 1+2\,w-8\,{w}^{2}-8\,{w}^{3} \right) \left( 1+4\,w\right)\cdot  w \nonumber \\
&&   \left( 1+2\,w-{w}^{2}-{w}^{3} \right)
  \left( 1-3\,w+{w}^{2} \right)  
\left( 1 +2\,w -4\,{w}^{2}\right)  \left( 1+w \right)  \nonumber \\
&&  \left( 16\,{w}^{8}-32\,{w}^{7}-17\,{w}^{6}+62\,{w}^{5}
-5\,{w}^{4}-35\,{w}^{3}+10\,{w}^{2}+3\,w-1 \right)  \nonumber \\
&& \left( 64\,{w}^{7}+96\,{w}^{6}+16\,{w}^{5}
-60\,{w}^{4}-21\,{w}^{3}+15\,{w}^{2}+8\,w+1 \right) \nonumber \\
&& \left( 128\,{w}^{5}-32\,{w}^{4}+48\,{w}^{3}+16\,{w}^{2}+4\,w-1 \right)
 \nonumber \\
&& \left( 4\,{w}^{5}+51\,{w}^{3}-21\,{w}^{4}-1+10\,w-35\,{w}^{2} \right)
 \nonumber \\
&& \left( 4\,{w}^{3}+7\,w-5\,{w}^{2}-1 \right)
\left( 4\,{w}^{4}+1+7\,w+26\,{w}^{2}+7\,{w}^{3} \right) \nonumber \\
&& \left( 4\,{w}^{4}+1+8\,w+20\,{w}^{2}+15\,{w}^{3} \right) \nonumber \\
&& \left( 1+12\,w+54\,{w}^{2}+112\,{w}^{3}+105\,{w}^{4}
+35\,{w}^{5}+4\,{w}^{6} \right)\, =\,\, 0.  \nonumber
\end{eqnarray}

We will see in the next section \ref{Landsin} that we missed the polynomials:
\begin{eqnarray}
 &&\left( 1+3\,w+4\,{w}^{2} \right)  \left( 1+4\,w+8\,{w}^{2} \right) 
 \left( 1-w \right)  \\
&&\qquad \qquad \qquad \left( 1+2\,w \right) 
\left( 1-w-3\,{w}^{2} +4\,{w}^{3} \right)  
\nonumber 
\end{eqnarray}

Note that we have not seen with the precision of these calculations
the occurence of the singularities of the $\, \Phi_H^{(3)}$.

With similar calculations using $1200$ terms for $\, \Phi_H^{(8)}$, the
following singularities are recognized:
\begin{eqnarray}
&& \left( 1-2\,w \right)  \left( 1+2\,w \right)  \left(1-2\,{w}^{2}
 \right)  \left( 1-4\,w \right)  \left( 1-4\,w+2\,{w}^{2} \right) 
 \left( 1+4\,w \right) \nonumber \\
&&  \left( 1+4\,w+2\,{w}^{2} \right)  \left( 8\,{w}^{2}-1 \right)
  \left( 3\,w-1 \right)  \left( 1-w \right) 
 \left( 1+w \right)  \left( 3\,w+1 \right) w \nonumber \\
&& \left( 1138\,{w}^{10}-1685\,{w}^{8}+
960\,{w}^{6}-242\,{w}^{4}+26\,{w}^{2}-1 \right) \nonumber \\
&&  \left( 32\,{w}^{4}-10\,{w}^{2}+1 \right)
 \left( 1312\,{w}^{6}-56\,{w}^{4}+30\,{w}^{2}-1 \right)
\nonumber \\
&&   \left( 10\,{w}^{2}-6\,w+1 \right) 
 \left( 4\,{w}^{3}-8\,{w}^{2}+6\,w-1 \right) \nonumber \\
&& \left( 5\,w-1 \right)  \left( 1+2\,{w}^{2}
 \right)  \left( 5\,w+1 \right) \nonumber \\
&&   \left( 10\,{w}^{2}+6\,w+1 \right) 
 \left( 4\,{w}^{3}+8\,{w}^{2}+6\,w+1 \right) \, =\, \,0.  \nonumber 
\end{eqnarray}

We will see in the next section \ref{Landsin} that we missed the polynomials:
\begin{eqnarray}
 \left( 1-3\,w +4\,{w}^{2}\right) 
 \left( 1+3\,w+4\,{w}^{2} \right) 
 \left( 1-10\,{w}^{2} +29\,{w}^{4}\right)  \nonumber
\end{eqnarray}

Note that the stabilized digits in these singularities can be as low
as two digits.

\section{Landau conditions and pinch singularities for
 $\Phi_D^{(n)}$ and integrals of $\,\prod y_j$.}
\label{bb}
Similarly to the integral representation (\ref{In-rep}) of $\, \Phi_H^{(n)}$, 
one has:
\begin{eqnarray}
\label{double}
&&\Phi_D^{(n)} \, = \, \, 
\int_{0}^{2\pi} {\frac{d\phi}{2\pi}} \int_{0}^{2\pi} \cdot  {\frac{d\psi}{2\pi}}
{{ \sqrt{(1\, -2w\, \cos\phi)^2 \, -4w^2 } } \over {D(\phi, \,\psi)  }} \\
&& \quad \quad \quad \quad \quad \quad \times
 {{ \sqrt{(1\, -2w\, \cos((n-1)\phi))^2 \, -4w^2 } }
 \over {D((n-1)\phi, \,(n-1)\psi)  }},
\nonumber 
\end{eqnarray}
and 
\begin{eqnarray}
\prod_{i=1}^{n} \, y_i \, = \, \,\,
 \int_{0}^{2\pi} \prod_{i=1}^{n} {\frac{d\phi_i}{2\pi}}\cdot  {\frac{d\psi_i}{2\pi}}
\cdot {{ 1} \over {D(\phi_i, \,\psi_i) }} 
\cdot  \delta\Bigl(\sum_{i=1}^{n}  \phi_i\Bigr).
\end{eqnarray}

For $\Phi_D^{(n)}$ the singularities of the 
associated ODEs are given as solutions of:
\begin{eqnarray}
&&D(\phi, \,\,\psi) \, = \, \, 0, 
\nonumber \\
&&D((n-1)\,\phi,\, \,(n-1)\,\psi) \, = \, \, 0, 
\nonumber \\
&&\alpha_1 \cdot \sin(\phi) \, + \, \alpha_2 \cdot \sin((n-1)\,\phi) \,= \, \, 0,
\qquad \quad \hbox{with} \quad \alpha_1, \, \, \alpha_2 \, \ne \, 0,
\nonumber \\
&&\alpha_1 \cdot \sin(\psi) \, 
+ \, \alpha_2 \cdot \sin((n-1)\,\psi) \,= \, \, 0
\end{eqnarray}
which are nothing less than the Landau conditions restricted to
pinch singularities of the singularity manifolds
 $\, D(\phi_i, \,\,\psi_i) \, = \, \, 0$.
For\footnote[8]{$\prod y_i$ or $\prod y_i^2$ integrand 
are similar as far as location of singularities
of the corresponding ODE is concerned.} $\prod_{i=1}^{n} y_i$
the singularities of the associated ODEs
can be written as the solutions of:
\begin{eqnarray}
&&D(\phi_i, \,\psi_i) \, = \, \, 0, \nonumber \\
&&\alpha_i \cdot \sin(\phi_i) \, 
- \, \alpha_n \cdot \sin(\phi_n) \,= \, \, 0,
 \quad i \, = \, \, 1, \,\cdots, \,  n-1, \quad
\hbox{with} \quad \alpha_i \, \ne \, 0 \, 
\nonumber \\
&&\alpha_i \cdot \sin(\psi_i) \, \,= \, \, 0, 
\qquad \qquad i \, = \, 1, \,\cdots,  \, N
\end{eqnarray}
which are also Landau conditions restricted to pinch singularities
of the singularity manifolds $\, D(\phi_i,\, \,\psi_i) \, = \, \, 0$.

\section{The singularities from Landau conditions}
\label{singLand}

In this Appendix, we give further details corresponding to
(\ref{famil1corps}), (\ref{famil2corps}) 
obtained from the Landau conditions:
\begin{eqnarray}
\label{elaneq1}
&& 1\,-2 w\cdot  \left( \cos(\phi_j) +\cos(\psi_j) \right) \,=\,\,\,0,
 \qquad \quad j=\,\,1,\,\cdots, \, n,  \\
\label{elaneq2}
&& \alpha_j\cdot  \sin(\phi_j)\, -\alpha_n \cdot \sin(\phi_n) \, =\,\,\,0,
 \qquad \quad \quad j=\,\,1,\,\cdots, \, n-1, \\
\label{elaneq3}
&& \alpha_j \cdot  \sin(\psi_j)\, -\alpha_n \cdot \sin(\psi_n) \,=\,\,\,0, 
 \qquad \quad \quad j=\,\,1,\,\cdots, \, n-1.
\end{eqnarray}
and:
\begin{eqnarray}
\label{angles3}
\sum_{j=1}^n \, \phi_j \,=\,\,0, \qquad \quad 
\sum_{j=1}^n \, \psi_j \,=\,\,0  \quad  \quad
   {\rm mod.} \, \, \,    2 \, \pi
\end{eqnarray}

We solve these equations for the values (zero or not) of 
$\,\sin(\phi_n)$ and $\,\sin(\psi_n)$.
For $\,\sin(\phi_n)\, =\,\sin(\psi_n)\, =\,0$, the case is simple and
gives $w\, =\,\pm 1/4$.

\subsection{The case $\, \sin(\phi_n)\, \ne\, 0$, $\sin(\psi_n)=\,0$}
\label{D1}
In this case, there are $k$ angles $\psi_j=\, \pi$
 and the remaining ones are $\psi_j=\, 0$.
By (\ref{angles}), the integer $k$ should be even, $k=\,2\,p$.
From (\ref{elaneq1}), we obtain and define\footnote[2]{Note that $\phi^{+}$ and $\phi^{-}$
(which correspond to $\, \psi_j\, = \, \pi$ and  $\, \psi_j\, = \, 0$ respectively)
are not on the same footing: indeed, the 
number of $\, \phi^{+}$ angles must be even, while the
number of $\phi^{-}$ angles depends on the parity of $\, n$.}
\begin{eqnarray}
\cos(\phi^{+}) \,=\, \,{1 \over 2w}\, +1, 
\qquad \quad \cos(\phi^{-}) \,\,=\,\, {1 \over 2w}\, -1.
\end{eqnarray}
One obtains $2p$ angles $\phi_j=\,\pm \phi^{+}$ and $\,n-2p\,$ angles
$\phi_j=\, \pm \phi^{-}$. The angles $\phi_j$ are then partitioned in
sets of $p_1$ angles $+\phi^{+}$, $\,(2p-p_1)\,$
 angles $-\phi^{+}$, $\, (n-2p-p_2)\,$
angles $+\phi^{-}$ and $\,p_2$ angles $-\phi^{-}$.
By (\ref{angles3}), one gets
 $\, (2p-2p_1)\cdot \phi^{+}=\, \, (n-2p-2p_2)\cdot\phi^{-}$.
Note that some manipulations on the indices lead to
$\cos\left(\vert 2p\vert \cdot \phi^{+} \right)=\, 
\cos\left(\vert n-2p-2k \vert \cdot \phi^{-} \right)$
and thus 
$\vert 2p \vert \cdot \phi^{+}=\, \pm \vert n-2p-2k \vert \cdot \phi^{-}$,
allowing us to write
\begin{eqnarray}
\label{famil1}
&& T_{2p} \left( 1/2w+1 \right)\, \, \,=\,\,\,\, \,
 T_{n-2p-2k} \left( 1/2w-1 \right), \\
&& 0 \,\,\le\,\, p\, \,\le \,\, [n/2], \quad \quad \quad 
 0\, \,\le\, \, k \,\,\le\,\, [n/2]\,-p, \nonumber
\end{eqnarray}
where $\, T_n(x)$ is the Chebyshev polynomial of the first kind.

One obtains the same results for the case $\, \sin(\phi_n)=0$ and
$\sin(\psi_n)\ne 0$.

\subsection{The case $\, \sin(\phi_n) \ne 0$, $\sin(\psi_n) \ne 0$}
In this case, by (\ref{elaneq2}), (\ref{elaneq3}),
 we have $\sin(\phi_j) \ne 0$ and
$\sin(\psi_j) \ne 0$. The equations (\ref{elaneq1}), (\ref{elaneq3}) become:
\begin{eqnarray}
\label{newlaneq1}
&& \cos(\psi_j) \, =\, \, \,  1\, -2 w \cdot  \cos(\phi_j), 
\qquad \quad \quad j\, =\, 1,\,\cdots,  \,  n,  \\
\label{newlaneq2}
&& \sin(\psi_j) \, =\, \,\,  \sin(\phi_j) \cdot
 {\frac{\sin(\psi_n)}{\sin(\phi_n)}}, \quad 
\qquad \quad j=\, 1,\,\cdots,  \,  n.
\end{eqnarray}
Squaring both sides of both equations and summing, one obtains
\begin{eqnarray}
\label{condphi}
\left( \cos(\phi_j)\,  -\cos(\phi_n) \right) \cdot
\left( \cos(\phi_j)\,  -\cos(\phi_0) \right) \,=\,\,\,  0,
\end{eqnarray}
where we have defined
\begin{eqnarray}
\label{cosphi0phin}
 \cos(\phi_0) \, =\,\, \, \,  
{\frac{4w\, -\cos(\phi_n)}{1\, -4w\,\cos(\phi_n) }}.
\end{eqnarray}

The angles $\, \phi_j$ are then partitioned into four sets $\pm \phi_0$ and
$\pm \phi_n$. Note that a similar condition (\ref{condphi}) occurs for
the angles $\psi_j$ which are partitioned likewise. Writing
(\ref{newlaneq1}), (\ref{newlaneq2}) 
for $j=\, 0$ and $j=\, n$ and with the
conditions (\ref{angles}), the equations become in terms
of Chebyshev polynomials\footnote[9]{Note that in equation (\ref{famil2})
one must realise that one takes 
the numerator of these rational expressions.}:
\begin{eqnarray}
\label{famil2}
&& T_{n_1}( z)\,  \, \, 
- T_{n_2} \Bigl(  {\frac{4w-z}{1-4w\,z }} \Bigr) \, = \,\,  0,  \\
&& T_{n_1} \left( {1 \over 2w}- z \right)\,\,\,  
 - T_{n_2} \Bigl( {1 \over 2w}- {\frac{4w-z}{1-4w\,z }} \Bigr) 
\,\, = \,\, \, 0, \nonumber \\
&& U_{n_2-1} (z) \cdot  
U_{n_1-1}\Bigl( {1 \over 2w}- {\frac{4w-z}{1-4w\,z }} \Bigr)\nonumber \\
&& \qquad \qquad 
- U_{n_2-1}\left( {1 \over 2w}- z \right)\cdot 
U_{n_1-1} \Bigl(  {\frac{4w-z}{1-4w\,z }} \Bigr)
\, \,=\,\,\, 0  \nonumber 
\end{eqnarray}
with
\begin{eqnarray}
&&  n_1 \,=\,\,  p, \qquad \qquad  n_2 \,=\,\,\,  n-p-2k,   \\
&&  0 \,\, \le\, \, p \,\, \le \,\,  n, \quad \quad \quad
0\, \, \le\,\,  k \, \, \le\,\,  [(n-p)/2]
\end{eqnarray}

At this step, some computational remarks are in order.
In the course of deriving (\ref{famil2}), some manipulations such as
dividing by a term have been done. Rigorously, the solutions that come
 from (\ref{famil2}) have to be checked against this point.
We have found, that as they are written, the formulas are ``safe'' 
from this perspective, except of the
 following. For $n=p/2$ (fixing $k=\, 0$ for convenience),
thus for $n$ even, the formulas (\ref{famil2}) give a common curve which
reads:
\begin{eqnarray}
w \, = \,\,\,\,  {1 \over 2 } \, {\frac{z}{1+z^2}}.
\end{eqnarray}
This relation comes from the condition $\cos(\phi_0)=\cos(\phi_n)$ in
(\ref{cosphi0phin}) which makes (\ref{condphi}) a perfect square.
We have checked that considering this condition at the outset, i.e.
(\ref{newlaneq1}), (\ref{newlaneq2}) yields no solution.

\subsection{Landau singularities}
\label{Landsin}
We can write the singularities obtained from (\ref{famil1}) as:
\begin{eqnarray}
&&n=3, \quad \quad \quad    \left( 1-4\,w \right)  \left( 1-w \right)
 \left( 1+3\,w+4\,{w}^{2} \right) \, = \, \, 0, \nonumber \\
&& n=4,\quad \quad \quad    \left( 1-16\,w^2 \right) 
 \left(1-4\,w^2 \right)  \, = \, \, 0, \nonumber \\
&&n=5, \quad \quad \quad    \left( 1-4\,w \right)   \left( 1-w \right)
\left( 1+3\,w+4\,{w}^{2} \right) 
 \left( 1-3\,w+{w}^{2} \right)  \nonumber \\
&& \quad \quad \times  \left( 1-7\,w+5\,{w}^{2}-4\,{w}^{3} \right)   
 \left( 1+8\,w+20\,{w}^{2}+15\,{w}^{3}+4\,{w}^{4} \right)
  \, = \, \, 0,  \nonumber \\
&& n=6, \quad  \quad  \quad   \left( 1-16\,w^2 \right) 
 \left( 1-4\,{w}^{2} \right) 
 \left( 1-{w}^{2} \right)  
  \left( 1-25\,{w}^{2} \right) \left( 1-9\,{w}^{2} \right) \nonumber \\
&&  \quad \quad   \times 
\left( 1\, +3w\, +4w^2 \right) \left(1-3w+4w^2\right) 
 \, = \, \, 0.  \nonumber
 \end{eqnarray}

The solutions of (\ref{famil2}) include
 some of the solutions of (\ref{famil1}).
We give in the following only those not occurring in  (\ref{famil1}):
\begin{eqnarray}
n=3, \quad \quad  && w \cdot  \left(1+4\,w \right) 
\left( 1+2\,w \right)  \, = \, \, 0,    \nonumber \\
n=4,\quad \quad  &&  w  \, = \, \, 0, \nonumber \\
n=5, \quad \quad  &&   w \cdot  (1+4\,w) 
\left( 1+w \right) \left( 1+2\,w \right)
 \left( 1+2\,w-4\,{w}^{2} \right)   \nonumber \\
&&  \times  \left( 1+4\,w+8\,{w}^{2} \right) 
\left( 1-w-3\,{w}^{2}+4\,{w}^{3} \right) \, = \, \, 0,  \nonumber \\
n=6, \quad \quad  && w \cdot  \left( 1-10\,{w}^{2}+29\,{w}^{4} \right) 
 \, = \, \, 0.  \nonumber 
\end{eqnarray}

All these singularities can be identified with the
 singularities occurring in the
linear ODE for $\, \Phi_H^{(n)}$, ($n=3, \cdots, 6$).

For $n=7$ and $n=8$, the solutions of (\ref{famil1}) and (\ref{famil2})
 can be identified with the singularities given in \ref{singphi7phi8} and
obtained in floating point form. They also give:
\begin{eqnarray}
n=7, \quad \quad  && 
 \left( 1+3\,w+4\,{w}^{2} \right)  \left( 1+4\,w+8\,{w}^{2} \right) 
 \left( 1-w \right)  \left( 1+2\,w \right)  \nonumber \\
&&  \times  \left( 1-w-3\,{w}^{2} +4\,{w}^{3} \right)   \,=\,0,  \nonumber \\
n=8, \quad \quad  && \left( 1-3\,w +4\,{w}^{2}\right) 
 \left( 1+3\,w+4\,{w}^{2} \right) 
 \left( 1-10\,{w}^{2} +29\,{w}^{4}\right) \,=\,0, \nonumber
\end{eqnarray}
which have not been found in the series
 with the currently available number of terms.

\section{Heegner numbers and other selected
 values of the modular $\, j$-function}
\label{d}

The nine Heegner numbers~\cite{jFunc} and their associated
 modular $\, j$-function $\, j(\tau)$,  yield the following conditions
in the variable $\, w$:
\begin{eqnarray}
j(1+i)\, = \,  (12)^3, \quad \quad \quad
\left( 1-8\,{w}^{2} \right)  
\left( 1-16\,{w}^{2}-8\,{w}^{4} \right) \, = \, \, 0,\nonumber 
\end{eqnarray}
\begin{eqnarray}
&&j(1+i\sqrt{2})\, = \,  (20)^3, \quad\quad \quad
 \left( 64\,{w}^{4}+16\,{w}^{2}-1 \right) \nonumber  \\
 && \qquad \qquad \quad  \times  
 \left( 64\,{w}^{8}+1792\,{w}^{6}-368\,{w}^{4}\, 
+32\,{w}^{2}-1 \right)\, = \, \, 0,\nonumber 
\end{eqnarray}
\begin{eqnarray}
j\Bigl( {{1+i\sqrt{3}} \over {2}} \Bigr)\, = \,  (0)^3, 
\quad \quad \quad 1 -16\,{w}^{2}+16\,{w}^{4} \, = \, \, 0, \nonumber 
\end{eqnarray}
\begin{eqnarray}
\label{1515}
&&j\Bigl(  {{1+i\sqrt{7}} \over {2}}\Bigr)\, = \,  (-15)^3,
 \quad\quad \quad\left( 1\, -31\,{w}^{2}   +256\,{w}^{4} \right)\,
  \left(1 \, -16\,{w}^{2} +{w}^{4}\right) \nonumber  \\
&& \quad \quad  \qquad \quad \quad \times   
\, (1\, +3\,w \, +4\,{w}^{2}) 
 \, (1\, -3\,w  +4\,{w}^{2}) 
 \, = \, \, 0, \nonumber 
\end{eqnarray}
\begin{eqnarray}
&& j\Bigl( {{1+i\sqrt{11}} \over {2}} \Bigr)\, = \,  (-32)^3,
 \quad \quad P_{3} \, \, = \, \, 
1 -48\,{w}^{2} +816\,{w}^{4} -5632\,{w}^{6} \nonumber   \\
&& \qquad \qquad \quad \qquad +45824\,{w}^{8}\, 
-536576\,{w}^{10}+4096\,{w}^{12} \, = \, \, 0, \nonumber 
\end{eqnarray}
and 
\begin{eqnarray}
&&j\Bigl( {{1+i\sqrt{d}} \over {2}} \Bigr)\, = \,  (-m)^3, 
\quad \quad \quad \quad P_{d} \, = \, \, 0 
\quad \quad \quad  \hbox{with:} \nonumber \\
&&\quad \quad \quad \quad  P_{d} \, = \, \,\,  
 P_{3} \,\, +\, N \cdot \left(1-16\,{w}^{2} \right)\cdot  {w}^{8}, 
  \nonumber 
\nonumber 
\end{eqnarray}
with the following values for the triplet $(d, \, m, \, N)$:
\begin{eqnarray}
&& (19, \, 96, \, 851968), \qquad (43, \, 960, \, 884703232),   \nonumber \\
&& (67, \,5280 , \, 147197919232), \qquad 
(163, \, 640320, \, 262537412640735232) \nonumber
\end{eqnarray}

Beyond Heegner numbers there are many other selected quadratic
values~\cite{selected,selected2} of $\, j$, for instance:
\begin{eqnarray}
\label{jquadra}
j\, = \,\,\, -4096 \cdot \left( 15+7\,\sqrt {5} \right)^{3}
 \,\, = \,\,\, j\Bigl( {{ 1+i \, \sqrt{35}} \over {2}} \Bigr)
\end{eqnarray}
Which is known~\cite{jFunc} to be one of the
 eighteen numbers having class number $\, h(-d)=2$,
and which corresponds to the quadratic relation 
$\, -134217728000+117964800\,j+{j}^{2}\, = \, \, 0$.
Recalling the expression of the modular $\, j$-function in term of the
variable $\, w$ 
\begin{eqnarray}
\label{jinw}
j \, = \, \, {\frac { \left( 1-16\,w^2 +16\,{w}^{4} \right)^{3}}
{ \left( 1-16\,{w}^{2} \right)\, {w}^{8} }}, 
\end{eqnarray}
this quadratic relation in $\, j$ 
becomes a quite involved polynomial expression 
that we have not seen emerging as singularities 
of (the linear ODE's of) our $\, n$-fold integrals.

\section{Landen transformations and the modular $\, j$-function}
\label{c}

In this Appendix the modular $j$-function (\ref{jfun}) will be seen, 
alternatively, as a function of the modulus $\, k$, and thus 
denoted $j[k]$, or as a function of the half period ratio $\, \tau$,
and  thus  denoted $\, j(\tau)$.
The modular function called the $j$-function when seen 
as a function of the modulus $\, k$ reads:
\begin{eqnarray}
\label{jfun2}
j[k] \, = \, \, \,256\cdot  
 {\frac { \left( 1-{k}^{2}+{k}^{4} \right)^{3}}{ k^{4} \cdot 
 \left( 1-k^{2} \right)^{2}}}.
\end{eqnarray}

Increasing the modulus by (\ref{landenA}), 
the modular function $\, j(k)$
becomes:
\begin{eqnarray}
\label{jFunc1}
j[k_1]\,=\,\, j_1[k] \, = \, \,\,
16 \cdot {\frac { \left( 1+14\,{k}^{2}
+{k}^{4} \right)^{3}}{ {k}^{2} \cdot
\left( 1-k^2 \right)^{4}}}.
\end{eqnarray}
Iterating this procedure once more one obtains:
\begin{eqnarray}
\label{jFunc2}
j_1[k_1]\,=\,\,j_2[k] \,\, = \, \,\,
4 \cdot {\frac { \left( {k}^{4}+60\,{k}^{3}\,
+134\,{k}^{2}+60\,k+1 \right)^{3}}
{k \cdot (1\, +k)^{2} \, (1-k)^{8}}}.
\end{eqnarray}

The decrease of the modulus by (\ref{landenD}) gives:
\begin{eqnarray}
\label{jFuncm1}
j[k_{-1}] \, =\,\, j_{-1}[k] 
  \,\, = \, \,\,
16 \cdot {\frac { \left( {k}^{4}-16\,{k}^{2}+16 \right)^{3}}{k^{8} 
\left( 1-k^{2} \right) }}.
\end{eqnarray}

The next iterations (the cube of
 (\ref{landenA}) and the square of
(\ref{landenD})) gives algebraic expressions for $\, j[k]$.

It is easy to get a {\em representation of  the Landen transformation
on the modular $\, j$-functions} by elimination
 of the modulus $\, k$ between
(\ref{jfun}) and (\ref{jFunc1}). One obtains the well-known
fundamental {\em modular curve}~\cite{Hanna,Myers}:
\begin{eqnarray}
\label{fondmodul}
&& \Gamma_1(j, \, j_1) \, = \, \,
 j^2 \cdot j_1^2 \, \, \, -(j+\, j_1) \cdot 
(j^2 \, +\, 1487 \, j\, j_1 \, +\, j_1^2) \nonumber \\
&&\quad \quad \quad \quad +3\cdot 15^3 \cdot
 (16 \, j^2  - \, 4027 \, j\, j_1 \, +\,16 \,  j_1^2) \\
&&\quad \quad \quad \quad \quad  -12 \cdot 30^6 \cdot (j\, + j_1) 
\,\, +8 \cdot 30^9 \,\, = \,\, \,\, 0.  \nonumber 
\end{eqnarray}

This algebraic curve is {\em symmetric} 
 in  $\, j$ and $\, j_1$. We will obtain
the same modular curve (\ref{fondmodul}) by elimination
 of the modulus $k$ between (\ref{jFunc1})
and (\ref{jFunc2}), or between (\ref{jfun2}) and (\ref{jFuncm1}).
The two modular functions $\, j$ and  $\, j_{1}$ are
invariant by the $\, SL(2, \, Z)$ modular group,
and, in particular, transformation $\, \tau \, \rightarrow \, 1/\tau$.
As a consequence, the transformation 
$\, \tau \, \rightarrow \, 2\cdot \tau$,
and its inverse $\, \tau \, \rightarrow \, \tau/2$, {\em have 
to be on the same footing} in  the modular curve 
representation (\ref{fondmodul})
for the Landen and Gauss transformations.

Similarly, one can easily find the (genus zero) 
modular curve $\, \Gamma_2$
obtained by the elimination of the 
modulus $k$ between (\ref{jfun2}) and (\ref{jFunc2}),
(or between (\ref{jFuncm1}) and (\ref{jFuncm1})), which
corresponds to the transformation
 $\, \tau \, \rightarrow \, 4\cdot \tau$
and, {\em at the same time}, to its 
inverse $\, \tau \, \rightarrow \,  \tau/4$.
This last algebraic curve is, of course,  {\em also a modular curve}. 

\subsection{Fixed points of these modular 
representations in terms of $\, j$-function}
\label{c7}

Transformations like $\, j \, \rightarrow \, j_1$, 
or $\, j \, \rightarrow \, j_2$,
corresponding to the previous modular curves, are not 
(one-to-one) mappings, they 
are called ``correspondence'' by Veselov~\cite{Veselov,Veselov2}. In order
 to look at the fixed points of
the Landen, Gauss transformations (or their iterates) 
seen as transformations on
{\em complex variables}, within the framework of (modular) representations
on the modular $\, j$-functions, we write, respectively, 
$\, \Gamma_1(j, j_1=j)= 0$ and $\, \Gamma_2(j, j_2=j)= 0$

The ``fixed points'' $\, \Gamma_1(j, j_1=j)=\,  0\, $ of
 the (modular) ``correspondence'' (\ref{fondmodul}), 
 are $\, j = \, j_1\, = \, \, (12)^3$
 or $\, (20)^3$ or $\, (-15)^3$. 

The ``fixed points'' $\, \Gamma_2(j, j_2=j)= 0$ of modular curve
 corresponding to the
square of the Landen transformation,  are
 $\, j = \, j_2\, = \, \, (66)^3$, or $\, 2 \cdot (30)^3$, 
or  $\, (-15)^3$ or the solutions\footnote[2]{This 
corresponds to a value of $\, j$ of class number
 $\, h(-d)=\, 2$, see (58) in~\cite{jFunc}.}
 of $\, j^2\, +191025 \cdot j\, -121287375 \, = \, 0$,
 namely: 
\begin{eqnarray}
\label{sqrt15}
j \, = \, \,-3^3 \cdot  \Bigl( {{1\, + \sqrt{5}} \over {2}}\Bigr)^2 \cdot 
(5\, + \, 4 \cdot \sqrt{5})^3 \, = \, \,
 j\Bigl(\tau \, = \,  \,  {{1\, + i\, \sqrt{15}} \over {2}}\Bigr)
\end{eqnarray}
and its Galois conjugate (change $\, \sqrt{5}$ into $\, -\sqrt{5}$).

\subsection{Alternative approach
 to fixed points of the Landen transformation
and its iterates}
\label{c8}

In order to get the ``fixed points'' of the Landen transformation,
  let us impose that (\ref{jfun2}) and (\ref{jFunc1})
 are actually equal, thus  $\, j[k] \, = \, j[k_1]$.
This yields the condition 
(already seen to correspond 
to the $\, \chi^{(3)}$-singularities 
$\, 1\, + 3\, w \, + 4\, w^2 \, = \, 0$):  
\begin{eqnarray}
\label{already}
\left( 4\,{k}^{2}+3\,k+1 \right) 
 \left( {k}^{2}+3\,k+4 \right) \, = \, \, 0
\end{eqnarray}
together with:
\begin{eqnarray}
\label{enplus}
&& \left( 4\,{k}^{2}-3\,k+1 \right)
 \left( {k}^{2}-3\,k+4 \right) \nonumber \\
&& \left( {k}^{2}+2\,k-1 \right)  \left( {k}^{2}-2\,k-1 \right)
 \left( 1+{k}^{2} \right)
 \, = \, \, 0.
\end{eqnarray}

The  first two polynomial conditions in (\ref{enplus}),
 $\,  (4\,{k}^{2}-3\,k+1) \, ( {k}^{2}-3\,k+4  ) \, = \, \, 0$,
correspond to the Heegner number associated with
the integer value $\, j \, = \, (-15)^3$.
The  next two  polynomial conditions in (\ref{enplus}),
 $ \, {k}^{2}\, \pm 2\,k-1 \, = \, 0$,
correspond to the Heegner number associated with
the integer value $\, j \, = \, (20)^3$.
The last
polynomial condition in (\ref{enplus}), $\, \, 1+{k}^{2} \, = \, \, 0$,
corresponds to the Heegner number associated with
the integer value $\, j \, = \, (12)^3$.

Similarly, in order to get the ``fixed points'' of
 the square of the Landen transformation,
  let us require that (\ref{jfun2}) and (\ref{jFunc2})
 are actually equal: $\, j[k] \, = \, j[k_2]$.
This yields the conditions (\ref{already})
 (fixed points of the Landen transformation)
together with:
\begin{eqnarray}
\label{equa2}
&&\left( {k}^{2}-6\,k+1 \right)  \left( 1+14\,{k}^{2}+{k}^{4} \right)
 \, = \, \,\, 0 \\
\label{equa3}
&&  \left( {k}^{4}-6\,{k}^{3}+17\,{k}^{2}+36\,k+16 \right)   \\
&& \qquad \qquad \times  \left( 16\,{k}^{4}+36\,{k}^{3}
+17\,{k}^{2}-6\,k+1 \right)  \, = \, \,\, 0. \nonumber 
\end{eqnarray}
In (\ref{equa2}) the condition 
$\,  1+14\,{k}^{2}+{k}^{4} \, = \, 0$ 
(or $\,  1-16w^2\, +256\, w^4 = \, 0$)
 corresponds to
$\, j\, = \, 2\, (30)^3 $ which is not a Heegner number
but actually corresponds to complex multiplication. 
The condition $\, k^2-6\,k+1\, = \, 0$  in (\ref{equa2}) 
(or $\, 1\, -32 \, w^2\, = \, 0$)
 corresponds to
$\, j\, = \, (66)^3 $ which is not a Heegner number either
but actually corresponds to complex multiplication.
Note that both polynomials under the Landen transformation (\ref{landenA})
give respectively $\, j\, = \, (0)^3 $ and
$\, j\, = \, (12)^3 $, i.e. Heegner numbers.
The last two (self-dual) conditions in (\ref{equa3}),
read in $\, w$
\begin{eqnarray}
&&1\, -9\,w\, +17\,{w}^{2}\, +24\,{w}^{3}+6\,{w}^{4} 
\, = \, \, 0,   \\
&&1\, +9\,w\, +17\,{w}^{2}\, -24\,{w}^{3}+6\,{w}^{4}
 \, = \, \, 0 \nonumber 
\end{eqnarray}
and yield as selected
 value~\cite{selected,selected2} of $\, j$, the quadratic
roots $\, -121287375\, +191025\, j+j^2\,= \, 0$, 
already given in (\ref{sqrt15}).

One more step can be performed writing the
 condition $\, j[k_{-1}] \, = \, j[k_2]$.
One gets the conditions:
\begin{eqnarray}
\left( {k}^{2}+3\,k+4 \right)^{2} \left( 4\,{k}^{2}-3\,k+1 \right)
 \left( {k}^{2}+2\,k-1 \right) 
 \left( {k}^{2}+1 \right) \, = \, \, 0 \nonumber
\end{eqnarray}
previously obtained and corresponding
 to $\, j \, = \, (-15)^3, \,20^3, \, 12^3$,
together with:  
\begin{eqnarray}
&&{k}^{6}-27\,{k}^{5}+363\,{k}^{4}+423\,{k}^{3}
-168\,{k}^{2}-144\,k+64 \, = \, 0,
\nonumber \\
&&{k}^{6}+17\,{k}^{5}+143\,{k}^{4}+203\,{k}^{3}
+52\,{k}^{2}+32\,k+64 \, = \, 0
\end{eqnarray}
corresponding, respectively, to the two cubic relations on $\, j$:
\begin{eqnarray}
\label{condi}
&&1566028350940383\, 
-58682638134\,j+39491307\,{j}^{2}+{j}^{3} \, = \, 0,
\nonumber \\
&& 12771880859375\, 
-5151296875\,j+3491750\,{j}^{2}+{j}^{3} \, = \, 0.
\end{eqnarray}
These conditions (\ref{condi}) yield quite involved
 polynomial expressions in the variable $\, w$ 
that we have not seen emerging as singularities 
of (the linear ODE's of) our $\, n$-fold integrals (or the 
$\, Y^{(n)}$ or $\, \Phi^{(n)}$ either). 

\section{Linear differential operators for the Sorokin integrals}
\label{G}

Recall the occurrence of zeta functions evaluated at integer
values in many $\, n$-fold integrals corresponding to
particle physics, field theory, ...
For instance, the following integral~\cite{Fischler, Fischler3}
is associated with $\, \zeta(3)$:
\begin{eqnarray}
\label{otherzeta3}
I_n (z) \,=\,
\int_{0}^{1}\,  du \, dv\, dw \cdot
  {{ (1-u)^n \, u^n \cdot (1-v)^n \, v^n \cdot (1-w)^n \, w^n }
 \over { (1\, -u\, v)^{n+1} \cdot  (z -u\, v\, w)^{n+1} }} 
\end{eqnarray}

From the series expansion of this holonomic $\, n$-fold integral,
we have obtained the corresponding order four Fuchsian
linear differential equation. 
On these linear differential operators the ``logarithmic'' 
nature of these integrals becomes clear.

The fully integrated series expansion of the triple integral
(\ref{otherzeta3}) is given by (where $x$ denotes $1/z$):
\begin{eqnarray}
&&
I_n(x) \,=\,\,\,
\sum_{i=0}^{\infty} x^{n+i+1} \cdot 
\frac{\Gamma^2(n+1) \cdot \Gamma^4(n+i+1)}{\Gamma(i+1)
 \cdot \Gamma^3(2+2n+i)} \times
\nonumber \\
&&\quad \quad \quad  \quad \quad _3F_2(n+1, n+i+1, n+i+1;2n+i+2, 2n+i+2;1).
\nonumber
\end{eqnarray}

The triple integral $\, I_n(x)$ is solution of the order four Fuchsian
linear differential operator ($Dx$ denotes $d/dx$)
\begin{eqnarray}
&&L_n\, =\, \,Dx^4 \,\,
+\frac{2 \, (3\, x\, -1)}{(x-1)\, x} \cdot Dx^3\nonumber \\ 
&& \qquad 
+\frac{\left(7\, x^2\, +(n^2+n-5)\, x\, 
-2\,n\,(n+1)\right)}{(x-1)^2 \,x^2 } \cdot Dx^2
\nonumber \\ 
&&\qquad + \frac{\left(x^2\,+2\,n\,(n+1)\right)}
{ (x-1)^2 \,x^3 } \cdot Dx \nonumber \\ 
&&\qquad +\frac{n\, (n+1) \cdot 
\left((n^2+n+1)\,x\,+(n-1)\,(n+2)\right)}{(x-1)^2\, x^4} 
\nonumber
\end{eqnarray}
which has the following factorization
\begin{eqnarray}
\label{operat}
&& L_n \, = \, \, \, 
\Bigl(Dx \, + \, {{d \ln(A_1)} \over {dx}}  \Bigr) 
\cdot \Bigl(Dx \, + \, {{d \ln(A_2)} \over {dx}}  \Bigr)  \\
&&\qquad 
 \quad\quad \quad\times  \Bigl(Dx \, + \, {{d \ln(A_3)} \over {dx}}  \Bigr) 
\cdot \Bigl(Dx \, + \, {{d \ln(A_4)} \over {dx}}  \Bigr) \nonumber
\end{eqnarray}
where:
\begin{eqnarray}
&&A_1\, = \, \,-(n-1)\cdot \ln(x) \, +2 \cdot \ln(x-1)  \, 
   +\ln(P_n), \nonumber \\
&&A_2\, = \, \, (n+1)\cdot \ln(x) \, -(n-1) \cdot \ln(x-1) 
 \,-\ln(P_n) \, +\ln(Q_n), \nonumber \\
&&A_3\, = \, \, -n \cdot \ln(x) \, +(n+1) \cdot \ln(x-1) 
 \,+\ln(P_n) \, -\ln(Q_n), \nonumber \\
&&A_4\, = \, \, n \cdot \ln(x) \, \,-\ln(P_n), \nonumber 
\end{eqnarray}
and where $\, P_n$ and $\, Q_n$ are polynomials in $\, x$ of degree $\, n$.
They are the polynomial solutions behaving as $ \cdots + x^n$ for
 a system of coupled differential equations ($P_n^{(m)}$ (resp. $Q_n^{(m)}$)
 denotes the $\, m$-th derivative of $P_n(x)$ (resp. $P_n(x)$)
with respect to $\, x$):
\begin{eqnarray}
&&  \left( x-1 \right)^{2} \cdot  {x}^{2} \cdot  P_n^{(4)} \nonumber \\
&& \,-2 \cdot 
 \left( 2\,(x -1)\cdot n\, -3\,x+1 \right)
\cdot \left( x-1 \right) \cdot  x \cdot   P_n^{(3)} \nonumber \\
&& +  \Bigl(\left( 2\,x-1 \right)  \left( 3\,x-4 \right) \cdot {n}^{2}\, 
- \left( 12\,{x}^{2}-13\,x +2\right) \cdot n\, + \left( 7\,x-5 \right)\, x \Bigr)
\cdot   P_n^{(2)} \nonumber \\
&& -  \left(2\, \left( 2\,x-3 \right)\cdot  {n}^{3}\, 
-2\, \left( 3\,x-1 \right)\cdot  {n}^{2}\, 
+2\, \left( 2\,x-1 \right)\cdot  n\, -x \right) \cdot P_n^{(1)} \, \nonumber \\
&&  +{n}^{4}\cdot P_n\,\, =\,\,\, \, 0, \nonumber
\end{eqnarray}
\begin{eqnarray}
&& -  \left( x-1 \right) \cdot  x \cdot  P_n\cdot   Q_n^{(2)}\,  \nonumber \\
&& + \left( 2 \, \left( x-1 \right)\cdot  x \cdot  P_n^{(1)}\, 
+ \left( 1-x+2\,n \right)\cdot   P_n \right) \cdot  Q_n^{(1)} \nonumber \\
&& - \left( 2\, \left( x-1 \right)\cdot  x \cdot P_n^{(2)}
-2\, \left((x\, -2)\,n \, -x\right)\cdot   P_n^{(1)}\, 
+ {n}^{2}\,P_n \right)\cdot   Q_n\, =\,\, 0. \nonumber
\end{eqnarray}

\vskip 0.8cm

\vskip .3cm

\vskip .3cm

\vskip .3cm

\vskip .3cm
\pagebreak

\end{document}